\documentclass[twocolumn,showpacs,prb,aps,superscriptaddress]{revtex4}
\usepackage{amsmath}
\usepackage{graphicx}
\usepackage{dcolumn}
\usepackage{color}
\newcommand{\be}{\begin{equation}}
\newcommand{\ee}{\end{equation}}
\newcommand{\bea}{\begin{eqnarray}}
\newcommand{\eea}{\end{eqnarray}}
\newcommand{\bs}{\begin{split}}
\newcommand{\bse}{\begin{subequations}}
\newcommand{\ese}{\end{subequations}}

\begin{document}

\title{Crystallography and Physical Properties of ${\rm\bf BaCo_2As_2}$, ${\rm\bf Ba_{0.94}K_{0.06}Co_2As_2}$ and ${\rm\bf Ba_{0.78}K_{0.22}Co_2As_2}$}

\author{V. K. Anand}
\altaffiliation{Present address: Helmholtz-Zentrum Berlin f\"ur Materialien und Energie, Hahn-Meitner Platz~1, D-14109 Berlin, \mbox{Germany}.}
\author{D. G. Quirinale}
\author{Y. Lee}
\author{B. N. Harmon}
\author{Y. Furukawa}
\affiliation {Ames Laboratory and Department of Physics and Astronomy, Iowa State University, Ames, Iowa 50011, USA}
\author{V.~V.~Ogloblichev }
\affiliation{Institute of Metal Physics, Ural Division of Russian Academy of Sciences, Ekaterinburg 620990, Russia}
\author{A. Huq}
\affiliation{Spallation Neutron Source, Oak Ridge National Laboratory, Oak Ridge, Tennessee 37890, USA}
\author{D.~L.~Abernathy}
\affiliation{Quantum Condensed Matter Division, Oak Ridge National Laboratory, Oak Ridge, Tennessee 37831, USA}
\author{P. W. Stephens}
\affiliation{Department of Physics and Astronomy, SUNY at Stony Brook, Stony Brook, New York 11794, USA}
\author{R. J. McQueeney}
\altaffiliation{Present address: Oak Ridge National Laboratory, Oak Ridge, Tennessee 37831, USA}
\author{A. Kreyssig}
\author{A. I. Goldman}
\author{D. C. Johnston}
\altaffiliation{johnston@ameslab.gov}
\affiliation {Ames Laboratory and Department of Physics and Astronomy, Iowa State University, Ames, Iowa 50011, USA}

\date{\today}

\begin{abstract}

The crystallographic and physical properties of polycrystalline and single crystal samples of ${\rm BaCo_2As_2}$ and K-doped Ba$_{1-x}$K$_x$Co$_2$As$_2$ $(x = 0.06, 0.22)$ are investigated by x-ray and neutron powder diffraction, magnetic susceptibility $\chi$, magnetization, heat capacity $C_{\rm p}$, $^{75}$As NMR and electrical resistivity $\rho$ measurements versus temperature~$T$\@. The crystals were grown using both Sn flux and CoAs self-flux, where the Sn-grown crystals contain 1.6--2.0~mol\% Sn.  All samples crystallize in the tetragonal ${\rm ThCr_2Si_2}$-type structure (space group $I4/mmm$).  For ${\rm BaCo_2As_2}$, powder neutron diffraction data show that the $c$~axis lattice parameter exhibits anomalous negative thermal expansion from 10~to~300~K, whereas the $a$~axis lattice parameter and the unit cell volume show normal positive thermal expansion over this $T$~range.  No transitions in ${\rm BaCo_2As_2}$ were found in this $T$ range from any of the measurements.  Below 40--50~K, we find $\rho\propto T^2$ indicating a Fermi liquid ground state.  A large density of states at the Fermi energy ${\cal D}(E_{\rm F}) \approx 18$~states/(eV\,f.u.) for both spin directions is found from low-$T$ $C_{\rm p}(T)$ measurements, whereas the band structure calculations give ${\cal D}(E_{\rm F}) = 8.23$~states/(eV\,f.u.).  The enhancement of the former value above the latter is inferred to arise from electron-electron correlations and the electron-phonon interaction.  The derived intrinsic $\chi(T)$ monotonically increases with decreasing~$T$, with anisotropy $\chi_{ab}>\chi_c$.  The $^{75}$As NMR shift data versus~$T$ have the same $T$ dependence as the $\chi(T)$ data, demonstrating that the derived $\chi(T)$ data are intrinsic.  The observed $^{75}$As nuclear spin dynamics rule out the presence of N\'eel-type antiferromagnetic electronic spin fluctuations, but are consistent with the presence of ferromagnetic and/or stripe-type antiferromagnetic spin fluctuations.  The crystals of ${\rm Ba_{0.78}K_{0.22}Co_2As_2}$ were grown in Sn~flux and show properties very similar to those of undoped ${\rm BaCo_2As_2}$.  On the other hand, the crystals from two batches of ${\rm Ba_{0.94}K_{0.06}Co_2As_2}$ grown in CoAs self-flux show evidence of weak ferromagnetism at $T \lesssim 10$~K with small ordered moments at 1.8~K of $\approx 0.007$ and $0.03~\mu_{\rm B}$ per formula unit, respectively.

\end{abstract}


\maketitle

\section{\label{Intro} Introduction}

The discovery of high-temperature superconductivity in the 122-type iron arsenides $A {\rm Fe_2As_2}$ ($A$ = Ca, Sr, Ba, Eu) at temperatures up to $T_{\rm c} = 38$~K upon chemical substitution or application of pressure motivated much research on this class of iron-based superconductors. \cite{Rotter2008a, Chen2008a, Sasmal2008, Wu2008, Sefat2008, Torikachvili2008, Ishida2009, Alireza2009, Johnston2010, Canfield2010, Mandrus2010, Damascelli2003, Lee2006} Extensive investigations have shown that the superconductivity can be achieved by partial chemical substitutions at the $A$, Fe or As sites.  In a search for unusual behaviors in other pure 122-type compounds that could potentially serve as parent compounds for high-$T_{\rm c}$ superconductivity or exhibit other novel ground states, we studied, e.g., ${\rm BaMn_2As_2}$, $A{\rm Pd_2As_2}$ ($A$ = Ca, Sr, and Ba) and $A{\rm Cu_{2-x}As_2}$ ($A$ = Ca, Sr). Antiferromagnetic (AFM) order was discovered in insulating ${\rm BaMn_2As_2}$ below a N\'eel temperature $T_{\rm N} = 625$~K arising from long-range ordering of the Mn spin $S=5/2$ local magnetic moments.\cite{Singh2009a, An2009, Singh2009b, Johnston2011}

On substituting K for Ba to form Ba$_{1-x}$K$_x{\rm Mn_2As_2}$, metallic hole-doped samples were produced for $x=1.6$\% to~40\%, where the doped holes evidently interact only weakly with the Mn local moments.\cite{Pandey2012,Lamsal2013}  At K-doping levels $x=19$ and 26\%, Bao et al.\ observed weak ferromagnetism (FM),\cite{Bao2012} which was later confirmed and studied in detail at a higher doping level of 40\%.\cite{Pandey2013a}  At the latter doping level, half-metallic FM below a Curie temperature of $\approx 100$~K was inferred to arise from the itinerant doped-hole carriers that coexists with the long-range AFM ordering of the Mn moments below $T_{\rm N} = 480$~K, where the ordered moments of the FM and AFM substructures are perpendicular to each other.\cite{Pandey2013a}

The compounds ${\rm SrCu_2As_2}$ (Ref.~\onlinecite{Anand2012a}) and ${\rm CaCu_{1.7}As_2}$ exhibit $sp$-band metallic behavior with a phase transition at 55~K in ${\rm CaCu_{1.7}As_2}$ (Refs.~\onlinecite{Cheng2012a, Anand2012b}) that may arise from ordering of the Cu vacancies.\cite{Anand2012b}  Bulk $s$-wave superconductivity below $T_{\rm c}=1.27$ and 0.92~K occurs in ${\rm CaPd_2As_2}$ and ${\rm SrPd_2As_2}$, respectively,\cite{Anand2013a} and below $T_{\rm c}= 1.69$ and~3.04~K in the related compounds ${\rm CaPd_2Ge_2}$ (Ref.~\onlinecite{Anand2014a}) and ${\rm SrPd_2Ge_2}$,\cite{Fujii2009} respectively.

Much work has been done to study and understand the superconducting $A$(Fe$_{2-x}$Co$_x)_2$As$_2$ systems with $x\ll1$,\cite{Johnston2010, Canfield2010, Leithe-Jasper2008, Nandi2010, Hu2011, Zhao2008, Ewings2011} but substantial research has now also been done on the $x=1$ metallic but nonsuperconducting endpoint compounds $A{\rm Co_2As_2}$ ($A$ = Ca, Sr, Ba).  These three compounds crystallize in the same body-centered tetragonal ${\rm ThCr_2Si_2}$-type structure (space group $I4/mmm$) as the $A{\rm Fe_2As_2}$ compounds do.\cite{Pfisterer1980}  A-type AFM order was observed in ${\rm CaCo_2As_2}$ below a sample-dependent N\'eel temperature 52--76~K,\cite{Cheng2012, Ying2012, Quirinale2013, Anand2014} in which the Co magnetic moments in an $ab$~plane align ferromagnetically along the $c$~axis, and have AFM alignment between adjacent Co planes.  Both the A-type AFM structure and the positive Weiss temperature in the Curie-Weiss law indicate that the dominant interactions in ${\rm CaCo_2As_2}$ are FM\@. The Co saturation moment from magnetization $M$ versus applied magnetic field $H$ isotherm measurements is only $\approx 0.3~\mu_{\rm B}$/Co, and a self-consistent modeling of the $M$ data in terms of the local-moment picture was not obtained,\cite{Anand2014} so the magnetism in ${\rm CaCo_2As_2}$ appears to be itinerant in nature. Furthermore, a significant vacancy concentration of $\approx7$\% on the Co sites was inferred both refinement of x-ray and neutron diffraction data as well as from wavelength-dispersive x-ray chemical analysis, so the composition of ``${\rm CaCo_2As_2}$'' is actually ${\rm CaCo_{1.86}As_2}$.\cite{Quirinale2013, Anand2014}  The presence of this large concentration of disordered vacancies on the Co site causes the $^{75}$As NMR spectrum to be very weak and diffuse.\cite{Furukawa2014}  The large range of $T_{\rm N}$ values given above that has  been observed in different studies of ``${\rm CaCo_2As_2}$'' may be due to a variable sample-dependent concentration of these vacancies.

The ${\rm SrCo_2As_2}$ and ${\rm BaCo_2As_2}$ compounds are enigmatic.  On the one hand, the large magnitudes of the magnetic susceptibility $\chi$ for both materials suggest within an itinerant magnetism picture that they should both be unstable to itinerant FM.\cite{Sefat2009, Pandey2013b} On the other hand, the $T$ dependence of $\chi$, the $H$ dependence of $M$ and other measurements on ${\rm SrCo_2As_2}$ and ${\rm BaCo_2As_2}$ show no evidence for any tendency towards FM ordering, or indeed for any type of static magnetic ordering above 2~K.\cite{Sefat2009, Pandey2013b}  In fact, the $\chi(T)$ for ${\rm SrCo_2As_2}$ exhibits a broad maximum at 115~K,\cite{Pandey2013b} which is characteristic of short-range dynamic AFM correlations in a low-dimensional spin system.\cite{Johnston1997}  Furthermore, inelastic neutron scattering measurements on ${\rm SrCo_2As_2}$ single crystals\cite{Jayasekara2013} revealed strong AFM fluctuations at the same ``stripe'' propagation vector as observed\cite{Johnston2010} in the paramagnetic and AFM states of the $A{\rm Fe_2As_2}$ parent compounds.  This result suggests that ${\rm SrCo_2As_2}$ is a plausible candidate for a high-$T_{\rm c}$ superconductor, so the lack of superconductivity in this material is a conundrum at present.  

Investigations of the physical properties of ${\rm BaCo_2As_2}$ self-flux-grown single crystals by Sefat et al.\cite{Sefat2009} revealed metallic behavior for both the $ab$-plane and $c$-axis electrical resistivity $\rho$ versus temperature $T$ from 2 to~300~K\@.  The low-$T$ $ab$-plane $\rho(T)$ followed a $T^2$ dependence, indicating a Fermi liquid ground state.  The $\chi(T)$ showed a large magnitude with a weak broad maximum at $\sim150$~K followed by a weak upturn at lower $T$.  No magnetic ordering was observed above 1.8~K\@. The low-$T$ heat capacity $C_{\rm p}(T)$ yielded a large Sommerfeld electronic heat capacity coefficent $\gamma$ indicating a large density of states at the Fermi energy ${\cal D}^\gamma(E_{\rm F}) = 17.4$~states/eV\,f.u.\ for both spin directions, where f.u.\ stands for formula unit.  Electronic structure calculations using the local density approximation (LDA) also gave a large ${\cal D}_{\rm band}(E_{\rm F}) = 8.5$~states/eV\,f.u.\ for both spin directions.\cite{Sefat2009}  Within an itinerant magnetism model the authors suggest that ${\rm BaCo_2As_2}$ should be ferromagnetic, but that long-range FM order is suppressed by quantum fluctuations due to proximity to a magnetic quantum critical point.\cite{Sefat2009}  

Two angle-resolved photoemission spectroscopy (ARPES) studies reveal that the Fermi surface of ${\rm BaCo_2As_2}$ exhibits  no obvious significant Fermi surface nesting,\cite{Xu2013,Dhaka2013} in agreement with band structure calculations.\cite{Sefat2009, Xu2013, Dhaka2013}  This situation contrasts with the clear nesting between the hole and electron pockets in the paramagnetic (PM) states of the $A {\rm Fe_2As_2}$ compounds that results in long-range itinerant spin-density wave AFM at temperatures $T_{\rm N}\lesssim 200$~K at the ``stripe'' wavevector spanning the electron and hole pockets.\cite{Johnston2010}  On the other hand, as noted above it was discovered that AFM fluctuations occur in the closely related compound ${\rm SrCo_2As_2}$ at the {\it same} stripe wavevector\cite{Jayasekara2013} even though there is no obvious Fermi surface nesting at that wavevector as deduced from ARPES data and band structure calculations for this compound.\cite{Pandey2013b}

In view of the above results on ${\rm CaCo_2As_2}$ and especially on ${\rm SrCo_2As_2}$, it is of great interest to confirm and extend the previous\cite{Sefat2009} physical property measurements on ${\rm BaCo_2As_2}$.  Furthermore, if ${\rm BaCo_2As_2}$ is indeed situated close to a magnetic quantum critical point as suggested,\cite{Sefat2009} one might expect the physical properties to change appreciably with doping.  We therefore further investigated the physical properties of undoped ${\rm BaCo_2As_2}$ and studied the effects of partial K-substitution for Ba in Ba$_{1-x}$K$_x$Co$_2$As$_2$ $(x = 0.06, 0.22)$. A polycrystalline sample of ${\rm BaCo_2As_2}$ was synthesized for neutron diffraction measurements.  Single crystals were synthesized using both Sn flux and CoAs self-flux and their properties were investigated by powder x-ray and neutron diffraction, $\rho(T)$,  $\chi(T)$, $M(H,T)$, $C_{\rm p}(T)$ and $^{75}$As nuclear magnetic resonance (NMR) measurements.  We also report electronic structure calculations of ${\rm BaCo_2As_2}$ and compare the results with the experimental data.

Our measurements indicate that ${\rm BaCo_2As_2}$ shows no phase transitions of any kind from 10 to 300~K\@.  However, the $c$-axis lattice parameter shows an anomalous negative thermal expansion over this temperature range, similar to but a factor of two smaller than our previous findings\cite{Pandey2013b} for ${\rm SrCo_2As_2}$.  We confirmed the $T^2$ dependence of the $ab$-plane $\rho(T)$ at low~$T$ and the Sommerfield coefficient in the low-$T$ $C_{\rm p}(T)$ found by Sefat et al.\cite{Sefat2009}  The $M(H)$ isotherms for our crystals grown out of Sn flux show that $M\propto H$ for temperatures from~2 to 300~K, with no evidence for saturation that might arise from FM correlations.  The anisotropy of $\chi$ is $\approx 20$\% for these crystals.  The temperature dependences of $\chi$ for crystals grown both with Sn flux and CoAs self-flux are similar to those reported in Fig.~1 of Ref.~\onlinecite{Sefat2009}.  However our $\chi(T)$ data, which are reproducible for different crystals,  disagree with those in Fig.~1 of Ref.~\onlinecite{Sefat2009} both in magnitude of $\chi$ and nature of the anisotropy.  Our calculations of the density of states versus energy ${\cal D}(E)$ show a sharp peak at the Fermi energy $E_{\rm F}$ that arises from the previously identified $d_{x^2-y^2}$ flat band at or near $E_{\rm F}$ in ${\rm SrCo_2As_2}$.\cite{Dhaka2013}  A large ${\cal D}(E_{\rm F})$ is also evident from the experimental Sommerfeld electronic heat capacity coefficient.

For a K-doped crystal grown in Sn flux with composition ${\rm Ba_{0.78}K_{0.22}Co_2As_2}$, we did not observe any significant difference in the properties from those of pure ${\rm BaCo_2As_2}$.  This result does not support the hypothesis\cite{Sefat2009} that ${\rm BaCo_2As_2}$ is close to a magnetic quantum critical point.  On the other hand, the magnetic measurements on two batches of CoAs self-flux grown crystals of Ba$_{1-x}$K$_x$Co$_2$As$_2$ with $x \approx 0.06$ reveal the presence of weak FM below 20~K with small ordered (saturation) moments of $\approx 0.007$ and 0.03~$\mu_{\rm B}$ per formula unit, respectively.

The experimental and theoretical details and crystallographic studies are presented in Secs.~\ref{ExpDetails} and ~\ref{Crystallography}, respectively. The physical properties are described in Secs.~\ref{BaCo2As2}, \ref{BaKCo2As2}, \ref{BaKCo2As2_CoAs} and~\ref{BaKCo2As2_Sn} for $x = 0$, 0.054, 0.065, and~0.22, respectively. The electronic density of states calculation for ${\rm BaCo_2As_2}\ (x=0)$ is presented in \ref{Sec:BaCo2As2_DOS}. A summary is given in Sec.~\ref{Conclusion}.

\section{\label{ExpDetails} Methods}

\begin{table*}
\caption{\label{tab:XRD1} Crystal and Rietveld refinement parameters obtained from powder x-ray diffraction data for crushed Ba$_{1-x}$K$_x$Co$_2$As$_2$ $(x = 0,~0.054,~0.065,~0.22)$ crystals and powder neutron diffraction data for a polycrystalline sample of ${\rm BaCo_2As_2}\ (x=0)$ crystallizing in the ${\rm ThCr_2Si_2}$-type body-centered tetragonal structure (space group $I4/mmm$). The atomic coordinates of Ba/K, Co and As atoms are (0,~0,~0), (0,~1/2,~1/4) and (0,~0,~$z_{\rm As}$), respectively. Also included are literature data for a polycrystalline sample with $x=0$ from Ref.~\onlinecite{Pfisterer1980} and for crushed single crystals with $x=0$ from Ref.~\onlinecite{Sefat2009}.}
\begin{ruledtabular}
\begin{tabular}{lcccccccc}
Composition & $x=0$ & $x=0$ & $x = 0$\footnotemark[1]  & $x=0$\footnotemark[2]	&$x=0$\footnotemark[1] & $x= 0.054$\footnotemark[1] & $x= 0.065$\footnotemark[1] & $x= 0.22$\footnotemark[1]  \\
   & Ref.~\onlinecite{Pfisterer1980} & Ref.~\onlinecite{Sefat2009} &  CoAs-flux & & Sn-flux & CoAs-flux & CoAs-flux & Sn-flux  \\
\hline
\underline{Crystal Data} \\
\hspace{0.8 cm} $a$ (\AA)     			&	3.958(5)	&	3.9537(1)		&  3.9522(4)  	& 3.95743(2)	& 3.9416(3) & 3.9427(6) & 3.9422(4) & 3.9191(6) \\
\hspace{0.8 cm} $c$ (\AA)     			&	12.67(2)	&	12.6524(6)	&  12.646(2)  	& 12.66956(9)	& 12.7073(8)& 12.694(2) & 12.696(2) & 12.818(2) \\
\hspace{0.8 cm} $c/a$         			&	3.201(1)	&	3.20014(7)	&  3.200(1)   	& 3.20146(4)	& 3.224(1)  & 3.220(1)  & 3.220(2)  & 3.271(2) \\
\hspace{0.8 cm} $V_{\rm cell}$ (\AA$^3$)	&	198.5(8)	&	197.78(2)		&  197.53(4) 	& 198.421(2)	& 197.43(2) & 197.32(5) & 197.31(3) & 196.88(6)\\
\hspace{0.8 cm} $z_{\rm As}$  			&	0.361	&				& 0.3509(3) 	& 0.35087(3)	& 0.3510(2) & 0.3516(4) & 0.3517(2) & 0.3514(4) \\
\underline{Refinement} \\
\hspace{0.63 cm}\underline{quality} \\
\hspace{0.8 cm}    $\chi^2$	            &		&		& 1.36 & 1.713 	& 	1.43 & 2.38 & 1.48 & 2.17 \\	
\hspace{0.8 cm}    $R_{\rm p}$ (\%)     &		&		& 14.3 & 10.10		& 14.6 & 17.2 & 14.8 & 17.5 \\
\hspace{0.8 cm}    $R_{\rm wp}$ (\%)    &		&		& 19.0 & 5.01		& 19.9 & 23.2 & 19.8 & 23.1 \\
\end{tabular}
\end{ruledtabular}
\footnotetext[1]{From laboratory-based powder x-ray diffraction data}
\footnotetext[2]{From synchrotron-based x-ray and neutron powder diffraction data that were refined together}
\end{table*}

A polycrystalline sample of ${\rm BaCo_2As_2}$ was synthesized by solid state reaction for the powder neutron diffraction measurements. The purity and sources of the elements used were Ba (99.99\%) from Sigma Aldrich, and Co (99.998\%) and As (99.99999\%) from Alfa Aesar.  Small pieces of Ba were mixed with Co and As powder in the stoichiometric ratio and pressed into a pellet that was then sealed in a tantalum tube under high purity Ar gas. The sealed tantalum tube was sealed inside an evacuated quartz tube and heated from room temperature to 590~$^\circ$C in 15~h, held for 15~h, heated to 620~$^\circ$C in 6~h, held for 15~h, heated to 750~$^\circ$C in 5~h, held for 10~h, heated to 780~$^\circ$C in 1~h, held for 3~h, heated to 800~$^\circ$C in 1/2~h, held there for 2~h and then furnace cooled. After this prereaction the sample was ground thoroughly and again pelleted, sealed in a tantalum tube under Ar that was then sealed in an evacuated quartz tube. The sample was heated to 1000~$^\circ$C in 16~h, held there for 2~h and then to 1100~$^\circ$C in 2~h, held for 48~h and then quenched into liquid nitrogen.

Single crystals of Ba$_{1-x}$K$_x$Co$_2$As$_2$ $(x = 0,~0.065, 0.22)$ were grown from solution using both Sn and prereacted CoAs as the flux.  The sources and purities of the elements used in addition to those given above were K~(99.95\%) and Sn~(99.999\%) from Alfa Aesar.  The chemical compositions of the crystals were determined using wavelength dispersive x-ray spectroscopy (WDS) analysis with a JEOL electron probe microanalyzer and energy dispersive x-ray spectroscopy (EDX) analysis using a JEOL scanning electron microscope. WDS analysis of self-flux grown ${\rm BaCo_2As_2}$ crystals revealed the desired 1\,:\,2\,:\,2 stoichiometry.

The Sn-flux growth of ${\rm BaCo_2As_2}$ was carried out with stoichiometric ${\rm BaCo_2As_2}$ and Sn in a 1:15 molar ratio placed in an alumina crucible that was sealed inside a partially argon-filled ($\approx 1/3$ atm pressure) silica tube that was then prereacted at 650~$^\circ$C for 12 h, heated to 900~$^\circ$C at a rate of 100 $^\circ$C/h and held for 5~h, then heated to 1100~$^\circ$C at a rate of 50~$^\circ$C/h, where it was held for 30~h, then slowly cooled to 600 $^\circ$C at a rate of 4~$^\circ$C/h\@. Thin shiny plate-like ${\rm BaCo_2As_2}$ crystals of typical size $3 \times 2 \times 0.2$~mm$^3$ were obtained by centrifuging the flux at the latter temperature.  WDS analysis of a Sn-flux grown ${\rm BaCo_2As_2}$ crystal revealed the presence of 1.6~mol\% Sn in the crystal. 

For the self-flux growth of ${\rm BaCo_2As_2}$ crystals, Ba and prereacted CoAs were taken in 1:4 molar ratio and placed in an alumina crucible that was sealed in a partially argon-filled ($\approx 1/4$ atm pressure) quartz tube. Crystal growth was carried out by heating to 1200 $^\circ$C at 60 $^\circ$C/h, holding for 5 h, heating to 1300 $^\circ$C at 50 $^\circ$C/h, holding for 1 h and then slowly cooling to 1180 $^\circ$C at 2 $^\circ$C/h. Shiny plate-like ${\rm BaCo_2As_2}$ crystals of typical size $4 \times 2.5 \times 1$~mm$^3$ were obtained upon decanting the flux with a centrifuge at 1180 $^\circ$C.

In order to grow Ba$_{1-x}$K$_x$Co$_2$As$_2$ crystals using Sn flux we started with stoichiometric Ba$_{0.5}$K$_{0.5}$Co$_2$As$_2$ and Sn in a 1:15 molar ratio placed in an alumina crucible that was sealed under high-purity Ar in a tantalum tube that was then sealed in a quartz tube with $\approx 1/3$ atm Ar pressure.  After prereaction at 650 $^\circ$C for 12~h it was heated to 1150~$^\circ$C following the same steps as above, where it was held for 35 h and then slowly cooled to 600~$^\circ$C at a rate of 3~$^\circ$C/h\@. Decanting the flux at this temperature using a centrifuge yielded plate-like crystals of typical size $2.5 \times 1.5 \times 0.1$~mm$^3$ which turned out to be ${\rm Ba_{0.78(1)}K_{0.22(1)}Co_2As_2}$ from EDX analyses.  This potassium concentration is much smaller than in the nominal starting ratio.  The presence of $\approx 2.0$~mol\% Sn was inferred from EDX analysis on a ${\rm Ba_{0.78}K_{0.22}Co_2As_2}$ crystal.  Sn is also incorporated into the structure of FeAs-based Ba$_{1-x}$K$_x{\rm Fe_2As_2}$ ($x=0$, 0.45) crystals grown in Sn flux.\cite{Ni2008} 

Self-flux growths of Ba$_{1-x}$K$_x{\rm Co_2As_2}$ crystals with $x>0$ were carried out using nominal starting compositions $x = 0.25, 0.5, 0.75$ and prereacted CoAs, again in a 1:4 molar ratio, placed in an alumina crucible, sealed in a tantalum tube under 1~atm Ar gas that was then sealed into a partially  Ar-filled ($\approx 1/4$ atm pressure) quartz tube. Crystals were obtained by heating to 1260~$^\circ$C following the above steps of self-flux growth, holding for 5 h and slow cooling to 1160~$^\circ$C at 2~$^\circ$C/h at which point the excess flux was removed by centrifuging. Shiny plate-like crystals of typical size $3 \times 2 \times 0.5$~mm$^3$ were obtained. In contrast to Sn-grown crystals where 22\% K doping was achieved, the K concentrations obtained from the EDX analyses of the self-flux grown K-doped ${\rm BaCo_2As_2}$ crystals were much smaller.  We obtained ${\rm Ba_{0.935(5)}K_{0.065(5)}Co_2As_2}$,  ${\rm Ba_{0.955(3)}K_{0.045(3)}Co_2As_2}$ and ${\rm Ba_{0.946(5)}K_{0.054(5)}Co_2As_2}$ for the starting growth concentrations $x = 0.25, 0.5, 0.75$, respectively. Since the difference in K~contents of these three compositions are very small, we measured the physical properties of crystals from the two batches with compositions ${\rm Ba_{0.935}K_{0.065}Co_2As_2}$ and ${\rm Ba_{0.946}K_{0.054}Co_2As_2}$.

The crystal structures of crushed crystals of ${\rm BaCo_2As_2}$, ${\rm Ba_{0.94}K_{0.06}Co_2As_2}$ and ${\rm Ba_{0.78}K_{0.22}Co_2As_2}$ were determined using laboratory-based powder x-ray diffraction (XRD) with Cu $K_\alpha$ radiation on a Rigaku Geigerflex \mbox{x-ray} diffractometer.  The magnetic measurements on the crystals were carried out using a Quantum Design, Inc., superconducting quantum interference device magnetic properties measurement system (MPMS). The contribution from the sample holder was subtracted from the measured magnetic moment except for the measurements in low magnetic fields ($H <0.1$~T). The heat capacity and electrical resistivity measurements were carried out using a Quantum Design, Inc., physical properties measurement system (PPMS). The heat capacity was measured by the relaxation method and the resistivity was measured by the standard four-probe ac technique using the heat capacity and ac transport options of the PPMS, respectively. A $^3$He attachment to the PPMS was used to measure the heat capacity down to 0.45~K.

High resolution x-ray powder diffraction patterns were collected at ambient temperature for crushed single crystals of ${\rm BaCo_2As_2}$ at the beamline X16C at the National Synchrotron Light Source.  The samples were loaded into a 1~mm diameter glass capillary along with Si powder which served as a standard for the refinement of the lattice parameters. The x-ray wavelength, 0.6995~\AA, was chosen using a Si(111) double monochromator.  The powder diffraction pattern was collected in the 5--45$^\circ$ 2$\theta$ range with a constant step size of 0.005$^\circ$, and a linearly varying counting time of 1--3 s/point. The incident beam intensity was monitored with an ion chamber and the diffracted radiation was measured with a NaI(Tl) scintillation detector. The axial and in-plane resolution of the diffractometer were set by slits and a Ge(111) analyzer crystal, respectively. 

Neutron powder diffraction data for a polycrystalline sample of ${\rm BaCo_2As_2}$ were collected at the Spallation Neutron Source (POWGEN) at Oak Ridge National Laboratory.\cite{Huq2011}  The sample was loaded into a vanadium sample can in the presence of He exchange gas to maintain thermal contact to the sample at low temperatures.  Data were collected as the sample temperature was increased from 10~K to 300~K in 15~K intervals using a central wavelength of 1.066~\AA\@.

Nuclear magnetic resonance (NMR) measurements were carried out for a CoAs self-flux grown ${\rm BaCo_2As_2}$ single crystal on the $^{75}$As nucleus ($I = 3/2,\ \gamma/2\pi = 7.2919$~MHz/T) using a homemade phase-coherent spin-echo pulse spectrometer.  $^{75}$As-NMR spectra were obtained by sweeping the magnetic field $H$ at a fixed frequency $f = 54.4$~MHz.  The magnetic field was applied parallel to either the crystal $c$~axis or the $ab$~plane.  The origin of the Knight shift $K = 0$ of the $^{75}$As nucleus was determined by $^{75}$As NMR measurements of GaAs.  The $^{75}$As nuclear spin-lattice relaxation rate (1/$T_{\rm 1}$) was measured with a saturation recovery method.

Electronic structure calculations were carried out for ${\rm BaCo_2As_2}$ using the full-potential linearized augmented plane-wave (FP-LAPW) method\cite{Blaha2001} with the local density approximation (LDA) \cite{Perdew1992} where we employed $R_{\rm MT}k_{\rm max} = 9.0$ with muffin tin (MT) radii $R_{\rm MT} = 2.1$~a.u.\ for all atoms in order to obtain self-consistent charge density.  828 {\bf k}-points were selected in the irreducible Brillouin zone and calculations were iterated to reach the total energy convergence criterion which was 0.01 mRy/cell. The body-centered tetragonal lattice parameters $a=b=3.958$~\AA, $c=12.670$~\AA\ from Ref.~\onlinecite{Pfisterer1980} were used.  The As atom was relaxed to get the theoretical $c$-axis position~$z_{\rm As} = 0.3441$.  This value is close to the experimental value $z_{\rm As} = 0.3509$ obtained in the following section.

\section{\label{Crystallography} Crystallography}

\subsubsection{Powder X-ray Diffraction Measurements on a Laboratory-Based Diffractometer}

\begin{figure}
\includegraphics[width=3in]{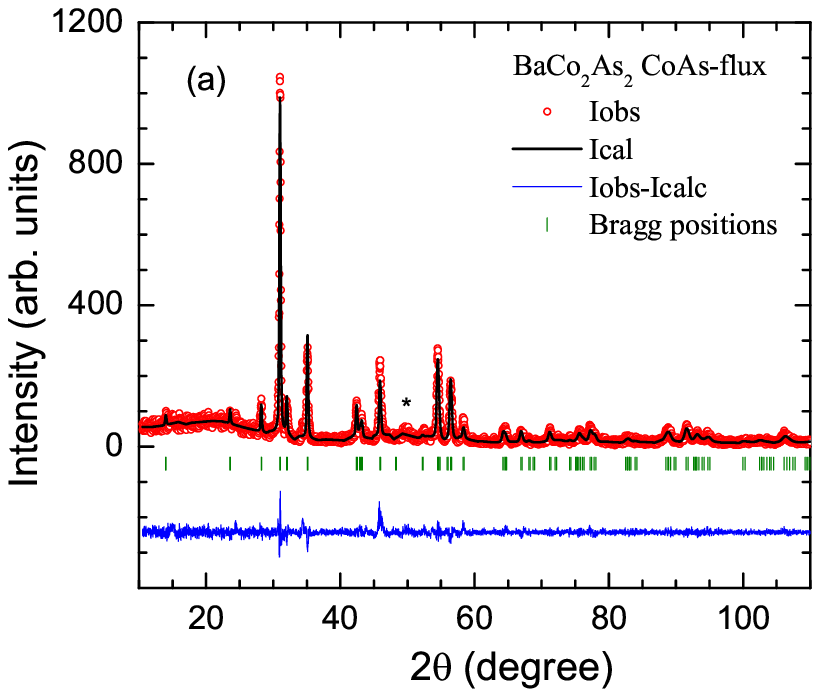}
\includegraphics[width=3in]{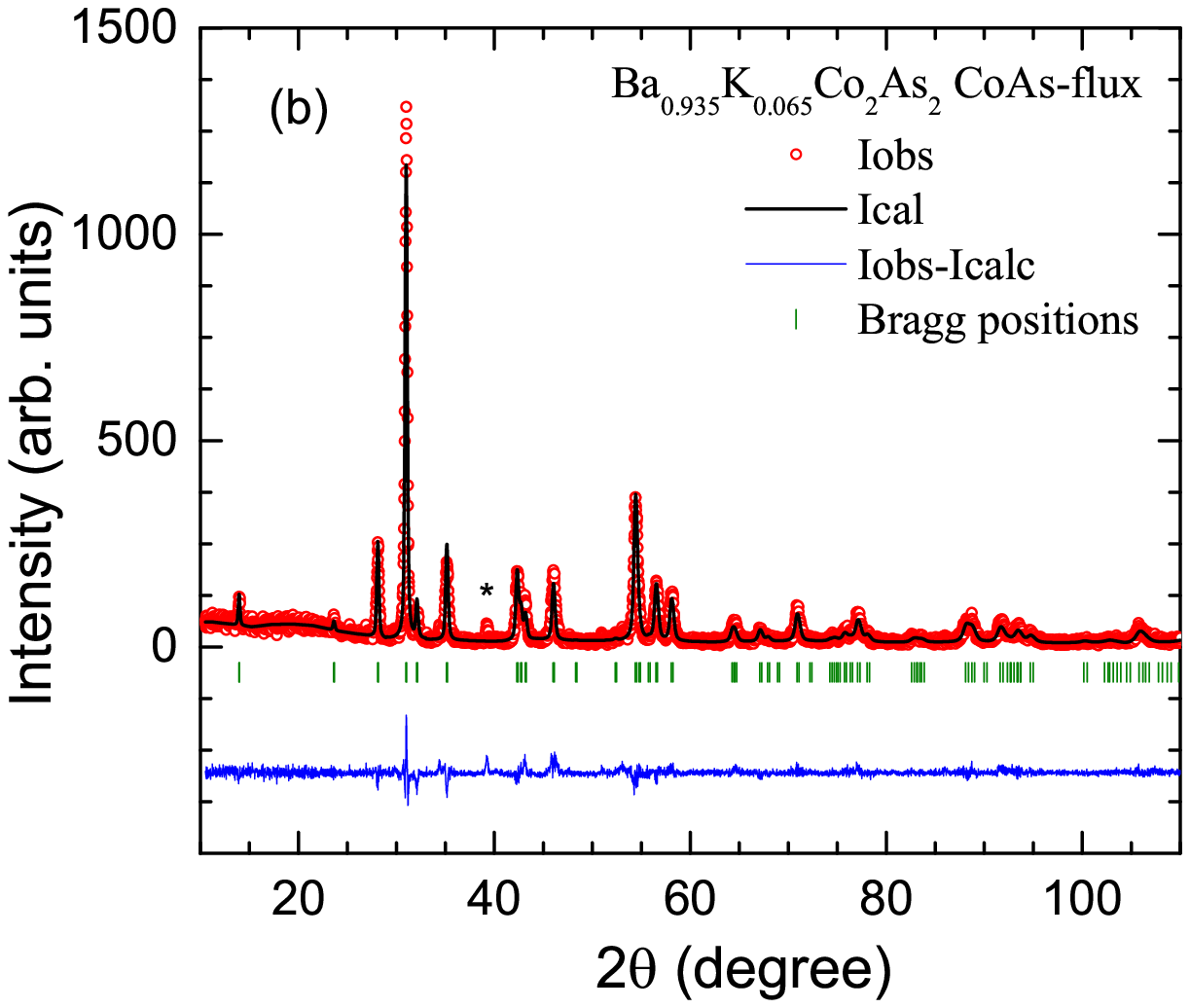}
\caption{(Color online) Powder x-ray diffraction patterns of (a) BaCo$_2$As$_2$ and (b) Ba$_{0.935}$K$_{0.065}$Co$_2$As$_2$ recorded at room temperature using a laboratory-based diffractometer. The solid lines through the experimental points are the Rietveld refinement profiles calculated for the ThCr$_2$Si$_2$-type body-centered tetragonal structure (space group $I4/mmm$). The short vertical bars mark the fitted Bragg peak positions. The lowermost curves represent the difference between the experimental and calculated intensities. The unindexed peaks marked with stars correspond to peaks from residual CoAs flux on the surface of the samples.}
\label{fig:BaCo2As2_XRD}
\end{figure}

The crystal structures of the compounds were determined from the analysis of room temperature powder XRD data collected on the crushed Ba$_{1-x}$K$_x$Co$_2$As$_2$ $(x = 0,~0.054,~0.065,~0.22)$ single crystals. The XRD data were analyzed by Rietveld structural refinement using the software {\tt FullProf} \cite{Rodriguez1993} confirming the single phase nature and ${\rm ThCr_2Si_2}$-type body-centered tetragonal structure ($I4/mmm$) of the compounds. The XRD patterns and Rietveld refinement profiles for two representative compositions $x=0$~and~0.065 are shown in Fig.~\ref{fig:BaCo2As2_XRD}. A few weak unindexed peaks are also observed in XRD patterns due to the residual CoAs flux on the surface of the crystals. During the refinement the thermal parameters $B$ were kept fixed to $B \equiv 0$ and the fractional occupancies of atoms were fixed to the stoichiometric values of unity as within the error bars the lattice parameters and the As $c$-axis position parameter $z_{\rm As}$ were insensitive to small changes in $B$ and occupancies of atoms. For K-doped compounds the fractional occupancies of Ba and K were kept fixed to the values as determined from the EDX composition analysis. The crystallographic and refinement parameters obtained are listed in Table~\ref{tab:XRD1}. Our crystal data for BaCo$_2$As$_2$ ($x=0)$ are in good agreement with the literature values\cite{Pfisterer1980, Sefat2009} that are also listed in Table~\ref{tab:XRD1}.

\subsubsection{Powder Diffraction Measurements on Synchrontron-Based X-ray and Neutron Diffractometers}

\begin{figure}
\includegraphics[width=3.in]{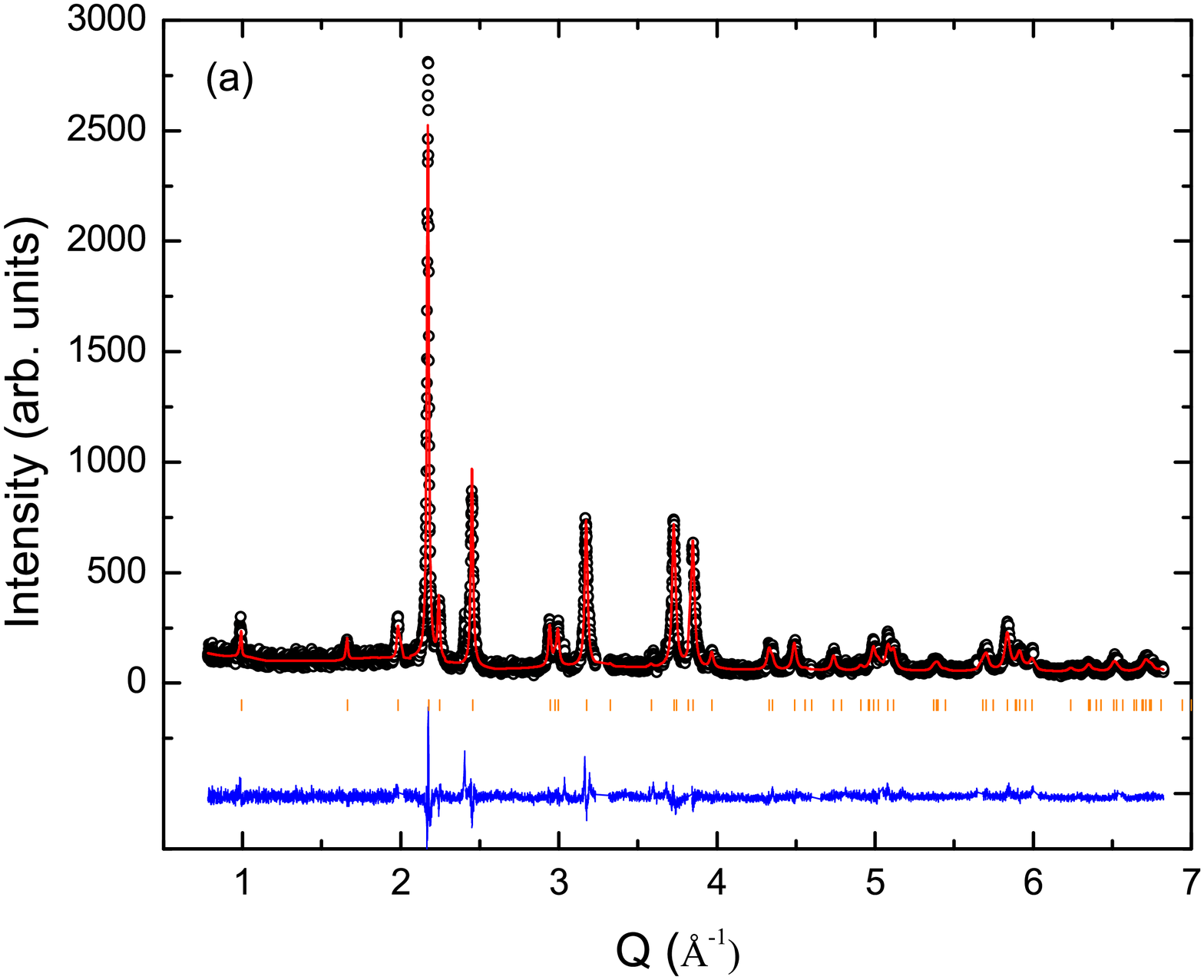}\vspace{0.1in}
\includegraphics[width=3.in]{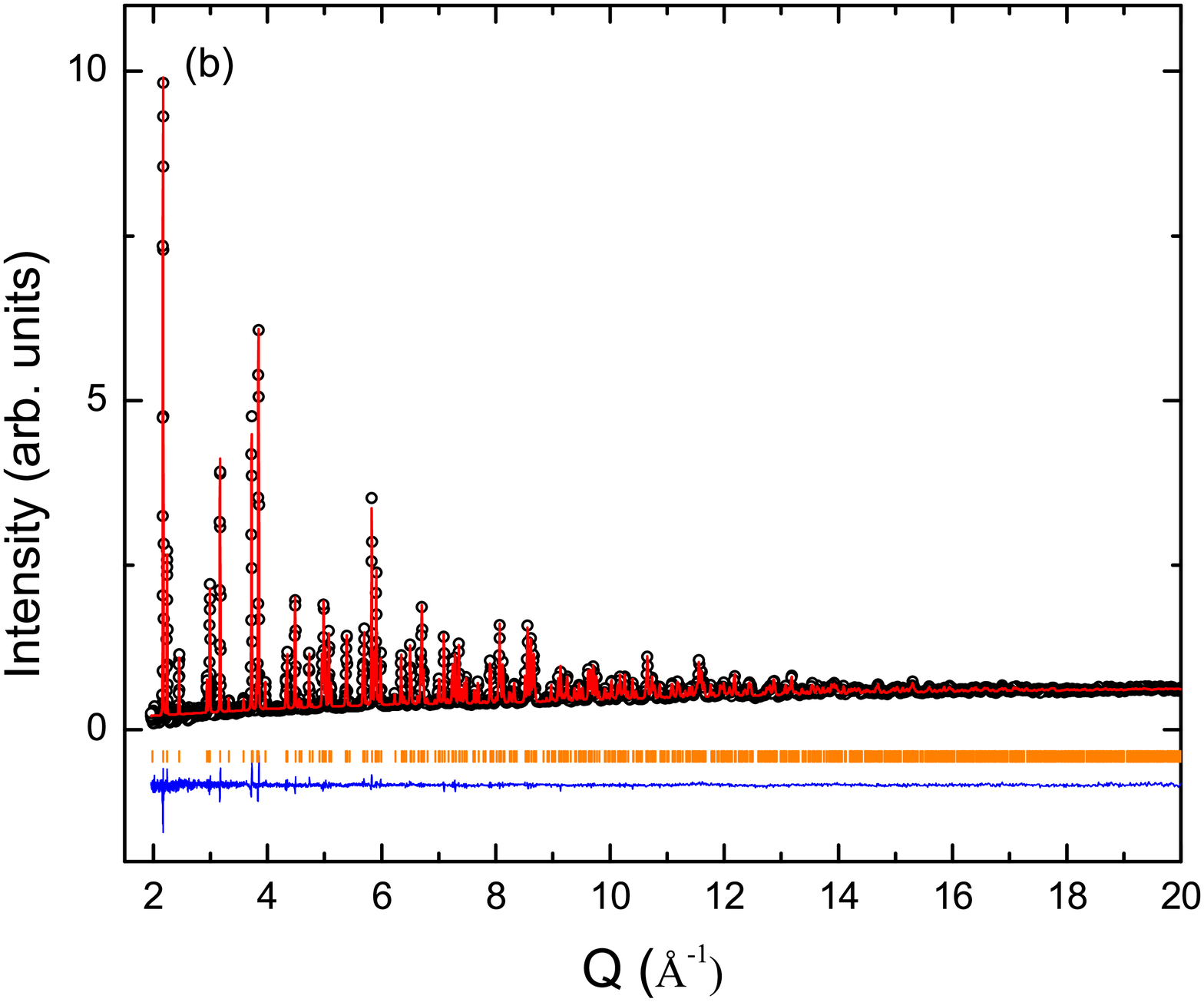}		              			
\caption{(Color online) (a) Synchrotron x-ray powder diffraction and (b)~neutron powder diffraction patters for crushed single crystals and a polycrystalline sample, respectively, of BaCo$_2$As$_2$ at 300~K\@. Gaps in the x-ray data correspond to regions where strong lines associated with the Si internal standard are found.  Shown are the measured data (open circles), the calculated profile fits from Rietveld co-refinement of the x-ray and neutron data (red lines), and the residuals (blue curve below fit).  The vertical orange tick marks correspond to the positions of the Bragg peaks of BaCo$_2$As$_2$.}
\label{BaCo2As2xrayneutron}
\end{figure}

\begin{figure}
\includegraphics[width=3.4in, viewport=-23 00 555 345,clip]{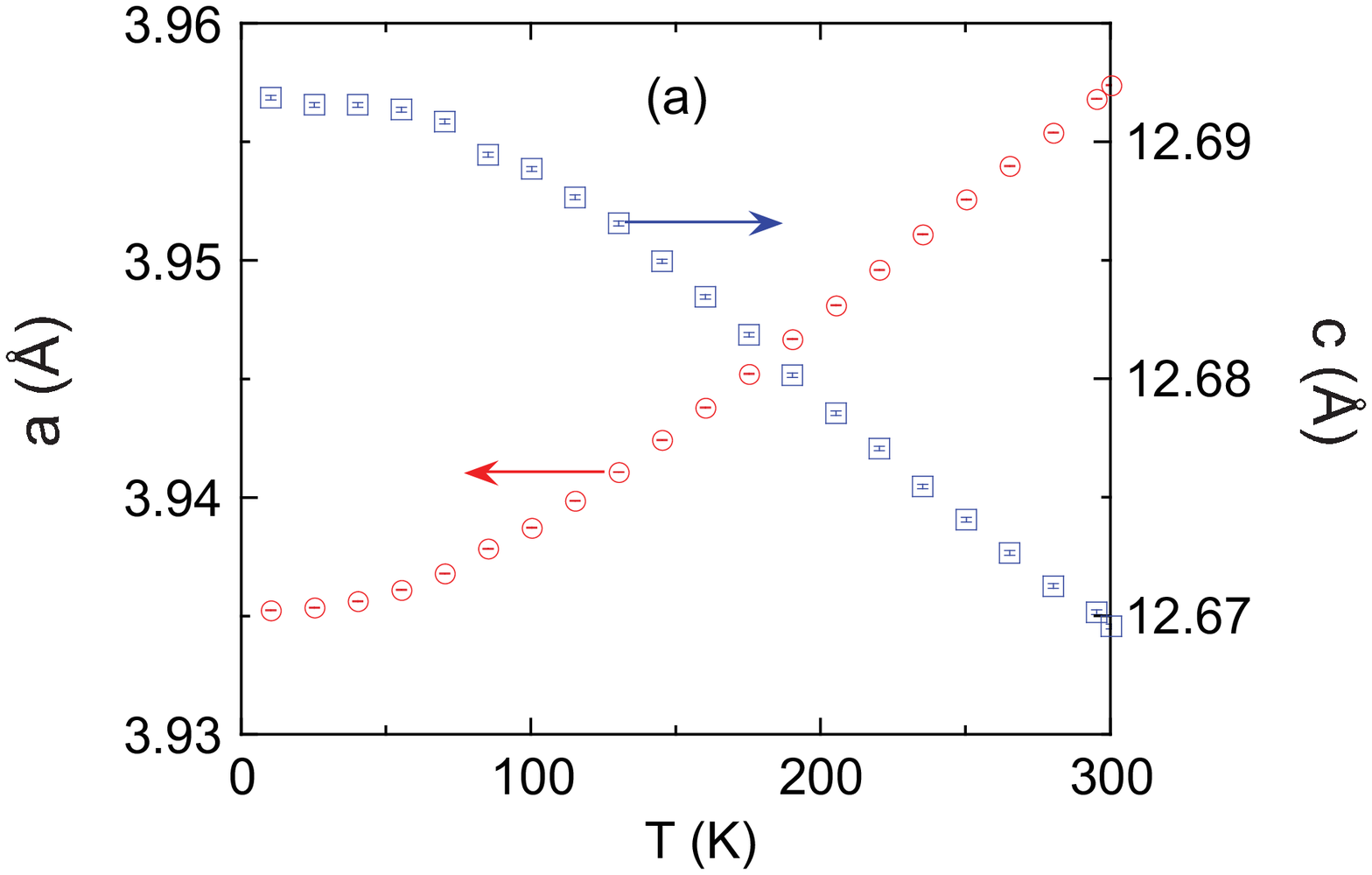}
\includegraphics[width=3.3in]{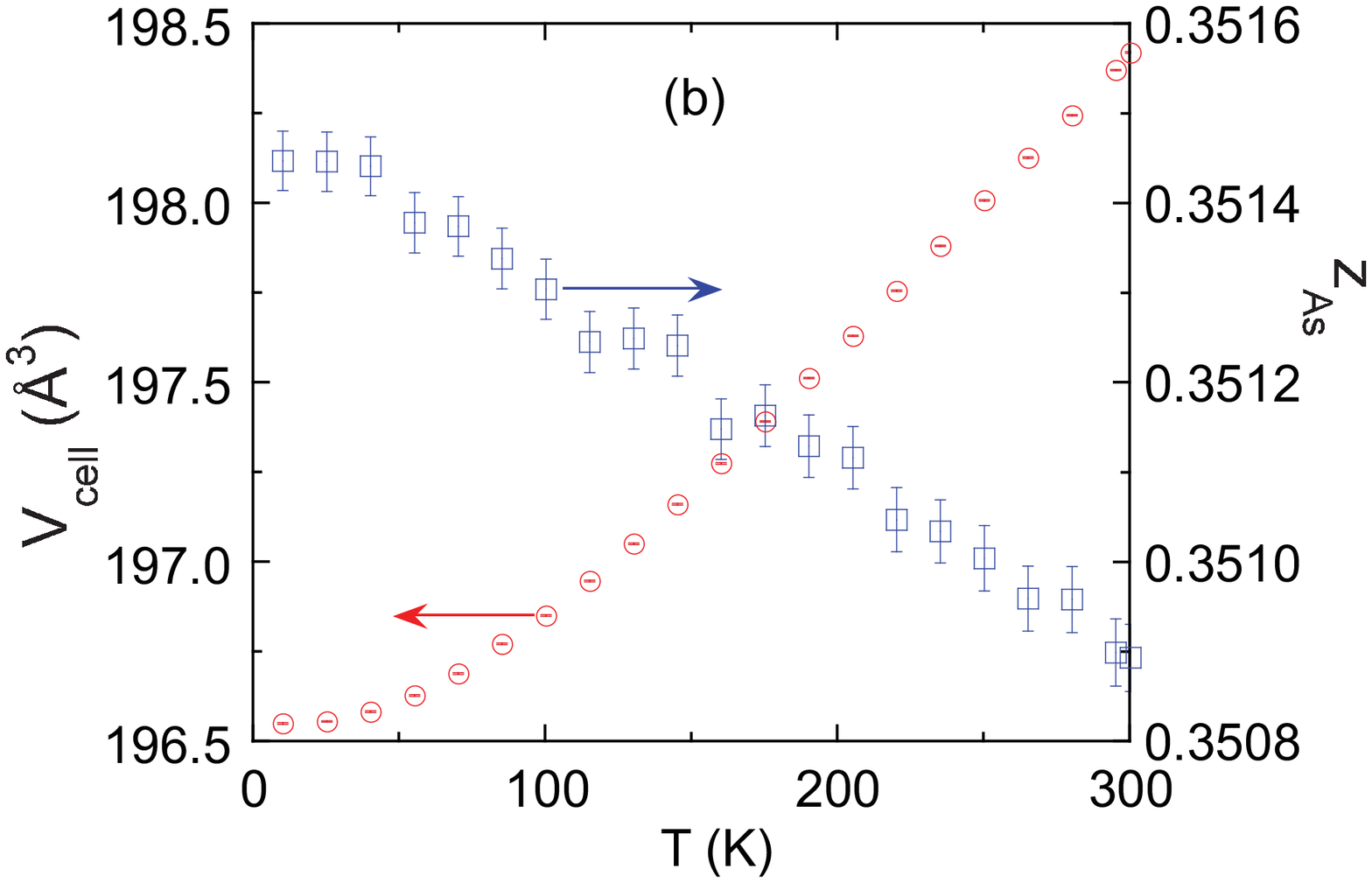}
\caption{(Color online) The temperature dependence of (a) the lattice parameters $a$ and $c$ and (b) the unit cell volume, $V_{\rm{cell}}$, and the As $c$-axis positional parameter, z$_{\rm{As}}$, for BaCo$_2$As$_2$ determined from refinements of the neutron powder diffraction data.  Error bars are plotted for each quantity.}
\label{Fig:BaCo2As2_a_c_zAs_vs_T}
\end{figure}

Rietveld co-refinements of the ambient-temperature x-ray and neutron powder diffraction patterns for BaCo$_2$As$_2$ were accomplished using the General Structure Analysis System package (GSAS)\cite{Larson2004} and the graphical user interface (EXPGUI).\cite{Toby2001} The neutron and x-ray powder data and refinements are shown in Fig.~\ref{BaCo2As2xrayneutron} and the results of the co-refinement are given in Table~\ref{tab:XRD1}. For the x-ray data, the strongest Si lines at low Q were excluded from the refinements. The principal impurity phase found in the sample of crushed single crystals grown in CoAs self-flux is adventitious CoAs which comprises approximately 8\% of the sample by weight.  A few very low intensity diffraction peaks in the x-ray powder pattern could not be identified conclusively with other impurity phases.

The refinements confirmed that BaCo$_2$As$_2$ crystallizes in the ${\rm ThCr_2Si_2}$-type body-centered tetragonal structure (space group $I4/mmm$).  The $c/a$ ratio of $\approx 3.2$ and the interlayer As--As distance $d_{\rm As-As} = (1-2z_{\rm As})c~\approx$~3.78~\AA\ confirm that BaCo$_2$As$_2$ has an uncollapsed tetragonal structure at ambient pressure, in contrast to CaCo$_{1.86}$As$_2$,\cite{Quirinale2013, Anand2014} which has a collapsed tetragonal structure arising from dimerization of the As atoms along the $c$~axis.\cite{Anand2012a}  The x-ray and neutron refinements (together and separately) show that, within experimental error, all atomic sites are fully occupied, again in contrast to the Co-deficiency observed for the Ca-based compound. \cite{Quirinale2013, Anand2014}

The temperature dependence of the $a$ and $c$~lattice parameters, as well as the unit cell volume and $z_{\rm{As}}$ determined from refinements of the neutron powder diffraction data are plotted in Fig.~\ref{Fig:BaCo2As2_a_c_zAs_vs_T}.  Similar to what was found for the closely related compound SrCo$_2$As$_2$, the $c$-axis thermal expansion coefficient $\alpha_c$ is negative from 10 to 300~K, whereas $\alpha_a$ (the $a$`axis thermal expansion coefficient) is positive over this same temperature range.  However, the change in the $c$~lattice parameter over this temperature range is roughly half of that noted previously for SrCo$_2$As$_2$ whereas the changes in the $a$~lattice parameters are comparable.  Nevertheless, as noted previously \cite{Pandey2013b} negative thermal expansion coefficients in the paramagnetic state are unusual.  The unit cell volume of BaCo$_2$As$_2$ smoothly decreases as temperature is lowered from 300~K to 10~K\@.  The $c/a$ ratio correspondingly increases by about 0.7\% on cooling, from 3.202 at 300~K to 3.225 at 10~K\@.  Finally, the arsenic positional parameter $z_{\rm As}$ increases with decreasing temperature over the measured range by roughly the same percentage as the $c$~lattice constant increases.

%\clearpage

\section{\label{BaCo2As2} Physical Properties of ${\rm\bf BaCo_2As_2}$}

\subsection{\label{Sec:BaCo2As2_Rho} Electrical Resistivity}

\begin{figure}
\includegraphics[width=3in]{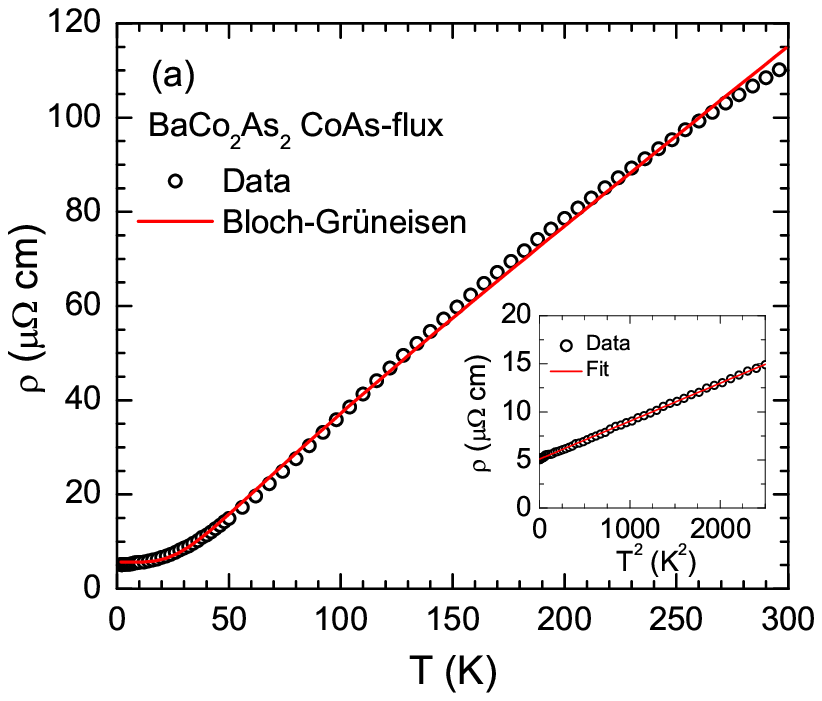}
\includegraphics[width=3in]{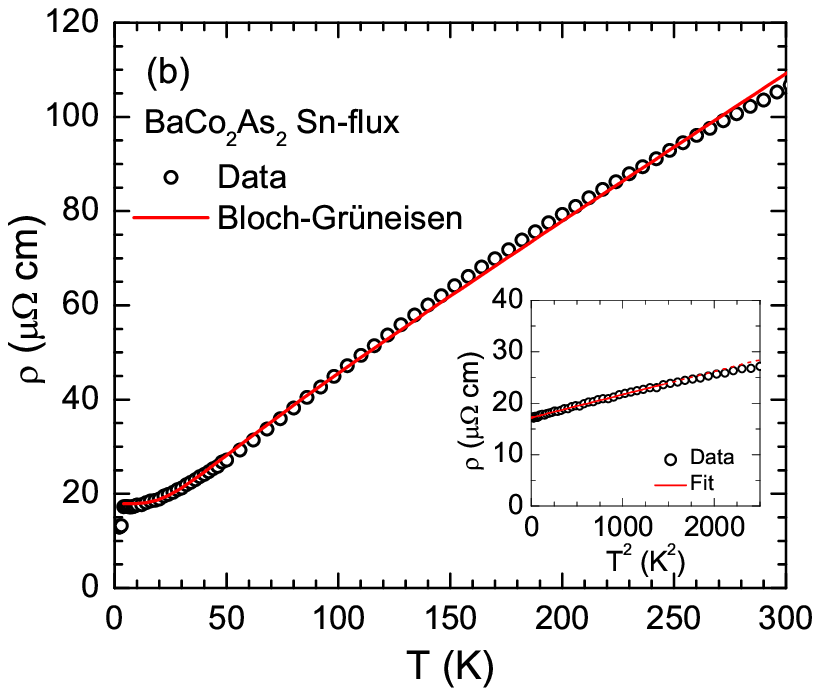}
\caption{(Color online) In-plane electrical resistivity $\rho$ of ${\rm BaCo_2As_2}$ single crystals grown using (a)~CoAs flux and (b)~Sn flux as a function of temperature $T$ measured in zero magnetic field. The red solid curves are fits by the Bloch-Gr\"{u}neisen model. Insets: $\rho$ vs $T^2$ plots below 50~K\@.  The straight lines are the fits by $\rho = \rho_0+AT^2$ for $1.8~{\rm K} \leq T\leq 50$~K in (a) and for $4.0~{\rm K} \leq T\leq 40$~K in (b).}
\label{fig:rho_BaCo2As2}
\end{figure}

The $T$ dependence of in-plane electrical resistivities $\rho$ of ${\rm BaCo_2As_2}$ crystals grown using CoAs and Sn flux are shown in Fig.~\ref{fig:rho_BaCo2As2}. The $\rho(T)$ data for both crystals exhibit metallic character.  The residual resistivity ratio is \mbox{RRR~$\equiv \rho(300\,{\rm K}) / \rho(1.8\,{\rm K}) = 21.8$} for the CoAs-grown crystal and 6.1 for the Sn-grown crystal.  The large RRR of the CoAs-grown crystal indicates good crystal quality.  The reason for the lower RRR of the Sn-grown crystal is likely due to the incorporation of Sn into the bulk of the crystal with a concentration of 1.6~mol\% as described in Sec.~\ref{ExpDetails}.  No signature of any phase transition is evident in the $\rho(T)$ data for either crystal except for a rapid drop in $\rho$ of the Sn-grown crystal at $T\approx 3.5$~K\@ due to the superconductivity of a small amount of adventitious Sn metal from the Sn flux. 

As shown in the insets of Figs.~\ref{fig:rho_BaCo2As2}(a) and (b) the $\rho(T)$ data exhibit a $T^2$ temperature dependence below 40--50~K\@. We analyzed the low-T data below 40--50~K by
\be
\rho(T) = \rho_0 + A T^2,
\label{eq:rho_T2}
\ee
where $A$ is a constant. The straight lines in the insets of Figs.~\ref{fig:rho_BaCo2As2}(a) and~(b) are the fits of $\rho(T)$ by Eq.~(\ref{eq:rho_T2}) for $1.8~{\rm K} \leq T\leq 50$~K in (a) and for $4.0~{\rm K} \leq T\leq 40$~K in~(b). The $T^2$ dependences of $\rho$ indicate a Fermi-liquid ground state in both crystals of ${\rm BaCo_2As_2}$. The Kadowaki-Woods ratio $R_{\rm KW} = A/\gamma^2$ for each crystal is given in Table~\ref{Tab:RhoFitParams}.  The value of $R_{\rm KW}$ for the CoAs-flux grown crystal is smaller than but comparable to the universal value of $R_{\rm KW} = A/\gamma^2 = 1.0 \times 10^{-5}~\mu \Omega\,{\rm cm/(mJ/mol\,K)^2}$ for strongly correlated and heavy fermion systems.\cite{Kadowaki1986}

\begin{table*}
\caption{\label{Tab:RhoFitParams} Parameters obtained from fits of the resistivities $\rho$ within the $ab$~plane of the listed single crystals by Eqs.~(\ref{eq:rho_T2}) and~(\ref{Eqs:BGModel}).  The quantity ${\cal R}$ is obtained from the fitted value of $\rho(\Theta_{\rm R})$ using Eq.~(\ref{eq:BG_R}).  The units of the Kadowaki-Woods ratio $R_{\rm KW} = A/\gamma^2$ are $10^{-5}~\mu \Omega\,{\rm cm/(mJ/mol\,K)^2}$.}
\begin{ruledtabular}
\begin{tabular}{lccccccc}
Compound 	& $\rho_0$ 		    &  $A$    										&  $R_{\rm KW}$ & $\rho_1$ 	&  $\Theta_{\rm R}$ 	& $\rho(\Theta_{\rm R})$  	& ${\cal R}$\\
		& ($\mu\Omega$\,cm)     & ($10^{-3}\,\frac{\mu \Omega\,{\rm cm}}{\rm K^2}$)	&   & ($\mu\Omega$\,cm) 		& (K)				& ($\mu\Omega$\,cm)			& ($\mu\Omega$\,cm)\\
\hline
${\rm BaCo_2As_2}$  (CoAs flux)   					& 5.10(2)      & 3.94(1)		& 0.20 	& 5.6(2)		&  174(5)   	& 61(2)  		&  64.4 \\	
${\rm BaCo_2As_2}$  (Sn flux)   					& 17.18(3)	& 4.50(5)  	& 0.28 	& 17.9(3)		&  140(6)   	& 41(2)  		&  43.3  \\	
${\rm Ba_{0.946}K_{0.054}Co_2As_2}$	(CoAs flux)    & 9.94(1)		& 3.31(1)		& 0.17 	& 10.4(1)		&  181(4)		& 54.6(1)		&  57.7\\
${\rm Ba_{0.935}K_{0.065}Co_2As_2}$  (CoAs flux)    	& 11.81(2)  	& 3.72(1)   	& 0.17 	& 12.2(2)  	&  164(3)		& 49.3(9) 	&  52.1 \\				
${\rm Ba_{0.78}K_{0.22}Co_2As_2}$ (Sn flux)         	& 22.07(3)	& 6.06(5) 	& 0.69 	& 22.8(2)		&  122(5)		& 38(2) 		&  40.1\\
\end{tabular}
\end{ruledtabular}
\end{table*}

The $\rho(T)$ data were further fitted over the full temperature range of the measurements by
\bse
\label{Eqs:BGModel}
\begin{equation}
\rho(T) = \rho_1 + \rho(T = \Theta_{\rm R}) \rho_{\rm {BG}}(T/\Theta_{\rm R}),
\label{eq:BG_fit}
\end{equation}
within the framework of Bloch-Gr\"{u}neisen (BG) model that describes the electrical resistivity $\rho_{\rm BG}(T)$ due to scattering of conduction electrons by longitudinal acoustic lattice vibrations and is given by \cite{Blatt1968}
\begin{equation}
\rho_{\rm {BG}}(T/\Theta _{\rm R})= 4 \mathcal{R} \left( \frac{T}{\Theta _{\rm R}}\right)^5 \int_0^{\Theta_{\rm{R}}/T}{\frac{x^5}{(e^x-1)(1-e^{-x})}dx},
\label{eq:Bloch-Gruneisen}
\end{equation}
where $\Theta_{\rm R}$ is the Debye temperature obtained from the resistivity data, $\mathcal{R}$ is a material-dependent prefactor and
\be
\rho(T/\Theta_{\rm R}=1) \approx 0.946\,464\,{\cal R}
\label{eq:BG_R}
\ee
\ese
is obtained from Eq.~(\ref{eq:Bloch-Gruneisen}).  Least-squares fits of the $\rho(T)$ data by Eqs.~(\ref{Eqs:BGModel}) for $1.8~{\rm K} \leq T\leq 300$~K and for $4.0~{\rm K} \leq T\leq 300$~K for CoAs- and Sn-flux grown ${\rm BaCo_2As_2}$ crystals are shown by the red curves in Figs.~\ref{fig:rho_BaCo2As2}(a) and (b), respectively, where we used the analytic Pad\'e approximant function \cite{Ryan2012} of $T/\Theta _{\rm R}$ in place of the integral on the right side of Eq.~(\ref{eq:Bloch-Gruneisen}). The fitting parameters are listed in Table~\ref{Tab:RhoFitParams} together with those of the other compounds studied in this paper. The value of $\mathcal{R}$ was obtained from the fitted value of $\rho(\Theta_{\rm{R}})$ using Eq.~(\ref{eq:BG_R}).  

%\clearpage

\subsection{\label{Sec:BaCo2As2_HC} Heat Capacity}

\begin{figure}
\includegraphics[width=3in]{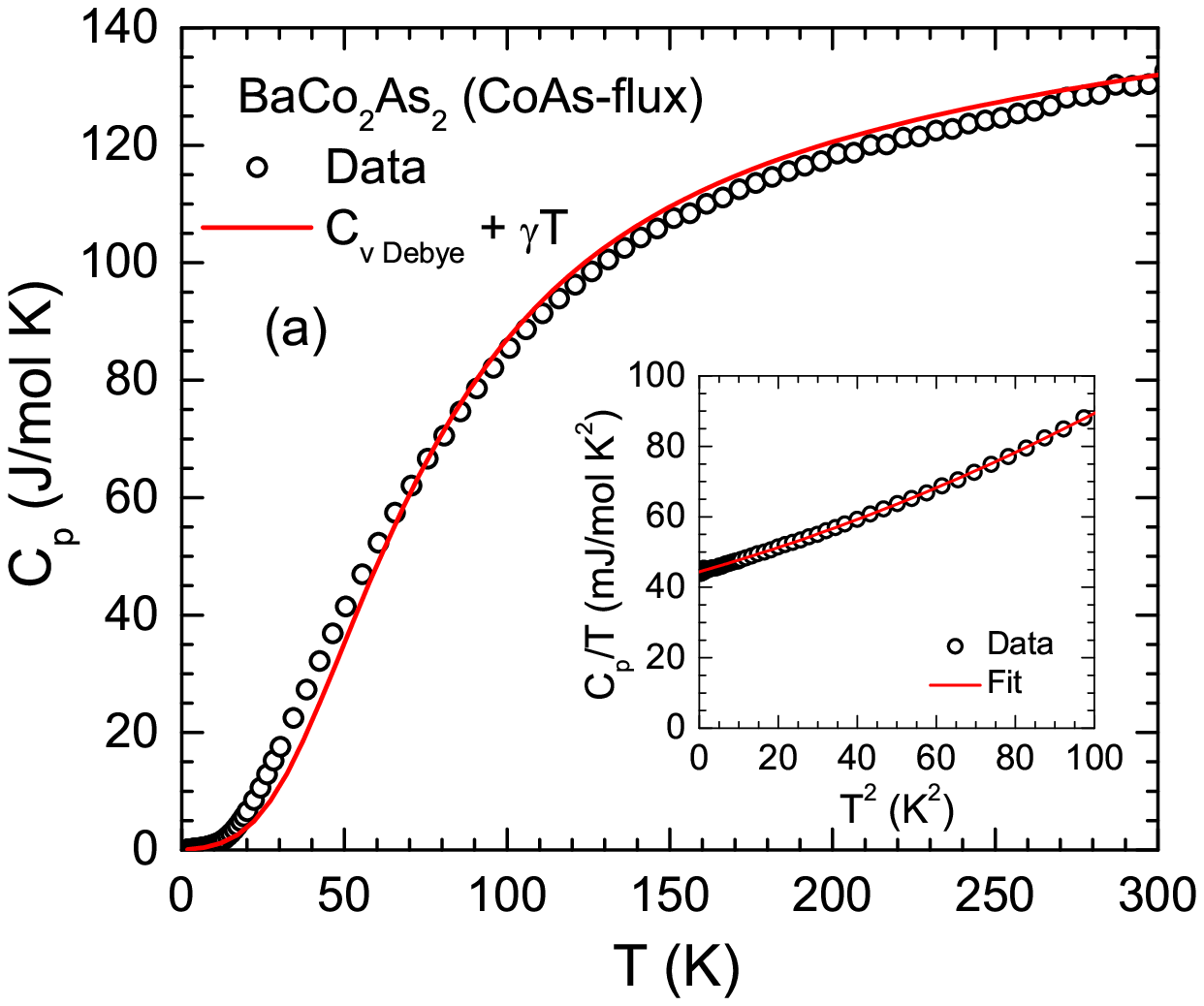}
\includegraphics[width=3in]{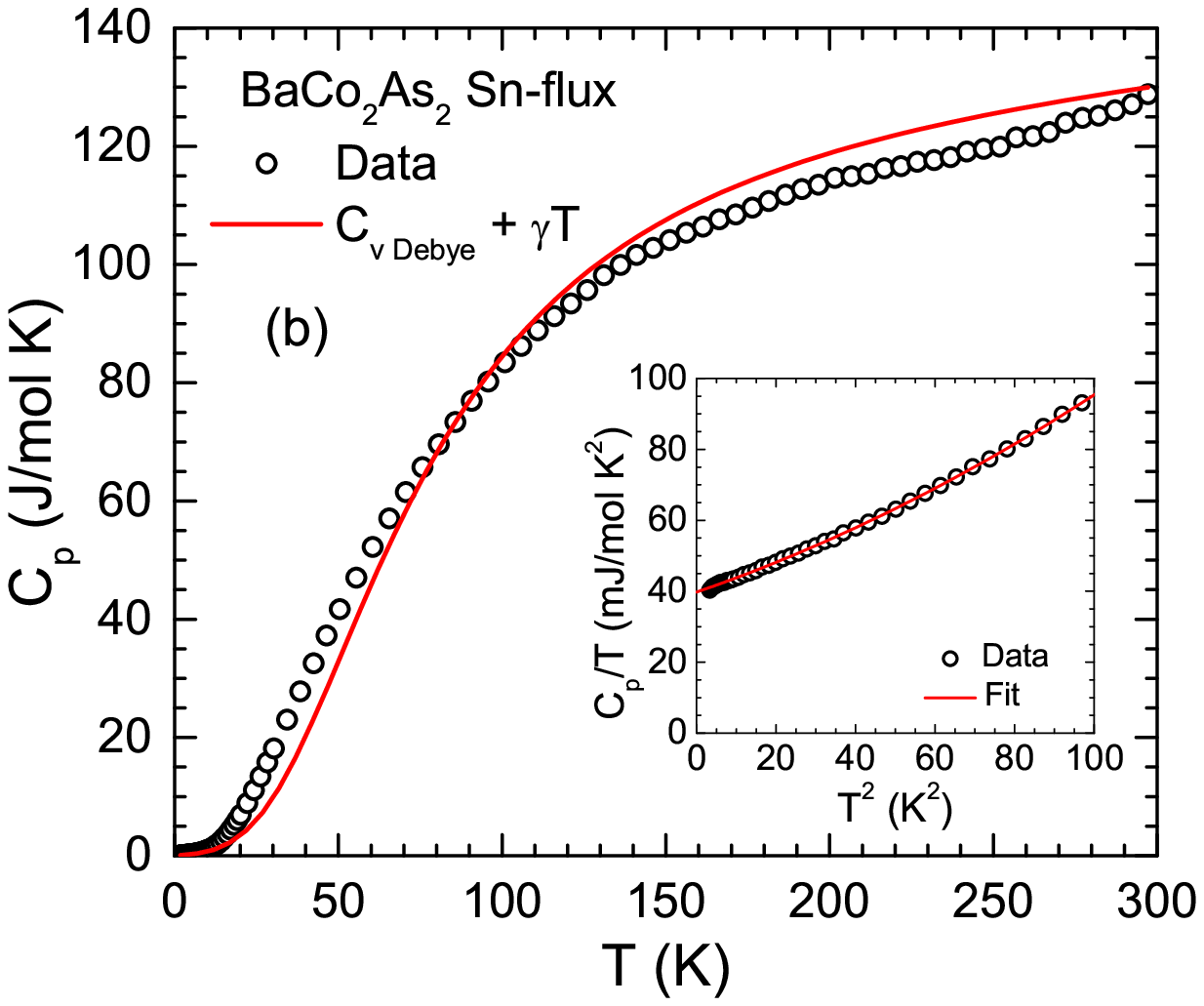}
\caption{(Color online) Heat capacity $C_{\rm p}$ of ${\rm BaCo_2As_2}$ single crystals grown using (a) CoAs~flux and (b) Sn~flux as a function of temperature $T$ measured in zero magnetic field for $1.8~{\rm K} \leq T\leq 300$~K\@. The solid curve is the sum of the contributions from the Debye lattice heat capacity $C_{\rm V\,Debye}(T)$ and electronic heat capacity $\gamma T$ according to Eq.~(\ref{eq:Debye_HC-fit}). Insets: $C_{\rm p}/T$ versus $T^2$ plot below 10~K, the red curves represent the fit of $C_{\rm p}/T$ data by Eq.~(\ref{eq:HC-LowT}) for $0.45~{\rm K} \leq T\leq 10$~K in (a) and $1.8~{\rm K} \leq T\leq 10$~K in (b).}
\label{fig:HC_BaCo2As2}
\end{figure}

The $T$ dependences of the heat capacity $C_{\rm p}$ of CoAs flux-grown and Sn flux-grown ${\rm BaCo_2As_2}$ crystals are shown in Figs.~\ref{fig:HC_BaCo2As2}(a) and \ref{fig:HC_BaCo2As2}(b), respectively.  No anomalies are observed between 0.45 and 300~K that would reflect the occurrence of phase transitions. The same value $C_{\rm p} (T= 300~{\rm K}) \approx 130$~J/mol\,K is observed for both the CoAs-flux and Sn-flux grown crystals, which is close to the expected classical Dulong-Petit value of $C_{\rm V} = 3nR = 15R$ = 124.7~J/mol\,K at constant volume, where $n=5$ is the number of atoms per f.u.\ and $R$ is the molar gas constant.\cite{Kittel2005, Gopal1966}

The low-$T$ $C_{\rm p}(T)$ data below 10~K are plotted as $C_{\rm p}/T$ versus $T^2$ are shown in the insets of Fig.~\ref{fig:HC_BaCo2As2}(a) and \ref{fig:HC_BaCo2As2}(b) for CoAs-flux and Sn-flux grown crystals, respectively. These data were fitted by
\be
\frac{C_{\rm p}(T)}{T} = \gamma + \beta T^2 + \delta T^4,
\label{eq:HC-LowT}
\ee
where $\gamma T$ is the Sommerfeld electronic heat capacity contribution and $\beta T^3+ \delta T^5$ is the low-$T$ lattice contribution. The $\gamma$, $\beta$ and $\delta$ parameters obtained from fits of the data in the inset of Fig.~\ref{fig:HC_BaCo2As2}(a) for $0.45~{\rm K} \leq T\leq 10$~K and the inset of Fig.~\ref{fig:HC_BaCo2As2}(b) for $1.8~{\rm K} \leq T\leq 10$~K by Eq.~(\ref{eq:HC-LowT}) are listed in Table~\ref{tab:tableHC}.  The coefficients $\gamma,~\beta~{\rm and}~\delta$ depend somewhat on the flux used for growth of the crystals and on the details of the associated heat treatment.    The fits are shown as the solid red curves in the respective insets.

The densities of states at the Fermi energy ${\cal D}(E_{\rm F})$ estimated from the $\gamma$ values using the relation \cite{Kittel2005}
\be
\gamma = \frac{\pi^2 k_{\rm B}^2}{3} {\cal D}^\gamma(E_{\rm F})
\label{eq:gamma_DOS}
\ee
are ${\cal D}^\gamma(E_{\rm F}) = 18.82(2)$ and 16.85(3)~states/eV\,f.u.\ for both spin directions for the CoAs~flux-grown and Sn~flux-grown crystals, respectively.  These $\gamma$ and ${\cal D}^\gamma(E_{\rm F})$ values for ${\rm BaCo_2As_2}$ are very large compared to those of the 122-type iron arsenide compounds.\cite{Chen2008b, Ronning2008, Rotundu2010}  The value $\gamma = 41.0$~mJ/mol\,K$^2$ obtained for ${\rm BaCo_2As_2}$ by Sefat et al.\cite{Sefat2009} is intermediate between our two values in Table~\ref{tab:tableHC}.  Large $\gamma$ values have also been observed in ${\rm CaCo_{1.86}As_2}$ (Ref.~\onlinecite{Anand2014}) and ${\rm SrCo_2As_2}$.\cite{Pandey2013b}

\begin{table*}
\caption{\label{tab:tableHC} The linear specific heat coefficients $\gamma$ and the coefficients $\beta $ and $\delta$ to the $T^3$ and $T^5$ terms in the low-$T$ heat capacity, respectively, and the density of states at the Fermi energy ${\cal D}(E_{\rm F})$ for both spin directions for Ba$_{1-x}$K$_x$Co$_2$As$_2$ $(x = 0,~0.05,~0.065,~0.22)$ single crystals. The Debye temperatures $\Theta_{\rm D}$ obtained for $0.45~{\rm K} \leq T \leq10$~K (low-$T$) and $0.45~{\rm K} \leq T \leq300$~K (all~$T$) fits to heat capacity data and the Debye temperatures $\Theta_{\rm R}$ obtained from fitting electrical resistivity data by the Bloch-Gr\"uneisen model for $1.8~{\rm K} \leq T \leq300$~K are also listed.}
\begin{ruledtabular}
\begin{tabular}{lccccccc}

Compound (crystal-growth flux)& $\gamma $  	& $\beta $  & $\delta$	&	${\cal D}^\gamma(E_{\rm F})$ 	&	$\Theta_{\rm D}$ (K)	&  $\Theta_{\rm D}$ (K)	& $\Theta_{\rm R}$ \\	
		 & (mJ/mol\,K$^2$)	& (mJ/mol\,K$^4$) &	($\mu$J/mol\,K$^6$) & (states/eV f.u.) & (low $T$)  &  (all~$T$)	& (K) \\
\hline
${\rm BaCo_2As_2}$  (CoAs flux)   					& 44.39(3)		&  0.319(3) 	& 1.32(4) 	& 18.82(2)      & 312(2)		& 301(3)  &  174(5) \\	
${\rm BaCo_2As_2}$  (Sn flux)   					& 39.75(7)		&  0.386(4) 	& 1.70(5) 	& 16.85(3)		& 293(2)  		& 311(4)  &  140(6) \\	
${\rm Ba_{0.946}K_{0.054}Co_2As_2}$	 (CoAs flux)    & 43.5(3)   	&  0.32(2) 		& 1.36(2)	& 18.4(2)	    & 312(7)		& 310(4)  &  181(4) \\
${\rm Ba_{0.935}K_{0.065}Co_2As_2}$  (CoAs flux)   	& 47.2(2)		&  0.271(7)		& 1.62(7)	& 20.0(1)		& 330(3)  		& 304(3)  &  164(3) \\		
${\rm Ba_{0.78}K_{0.22}Co_2As_2}$ (Sn flux)         & 29.7(3)		&  0.378(9)		& 1.44(8) 	& 12.6(2)		& 295(3)		&         &  122(5) \\
\end{tabular}
\end{ruledtabular}
\end{table*}

The Debye temperature $\Theta_{\rm D}$ was estimated from $\beta$ using the relation \cite{Kittel2005}
\begin{equation}
\Theta_{\rm D} = \left( \frac{12 \pi^{4} n R}{5 \beta} \right)^{1/3},
 \label{eq:Debye-Temp}
\end{equation}
yielding $\Theta_{\rm D} = 312(2)$~K and 293(3)~K for CoAs~flux-grown and Sn~flux-grown crystals, respectively.

An analysis of the $C_{\rm p}(T)$ data over the whole measured $T$~range gives another estimate of $\Theta_{\rm D}$. In this case, the $C_{\rm p}(T)$ data were fitted by
\bse
\label{Eqs:AllTCpFit}
\begin{equation}
C_{\rm p}(T) = \gamma T + n C_{\rm{V\,Debye}}(T),
\label{eq:Debye_HC-fit}
\end{equation}
where\cite{Gopal1966}
\begin{equation}
C_{\rm{V\,Debye}}(T) = 9 R \left( \frac{T}{\Theta_{\rm{D}}} \right)^3 {\int_0^{\Theta_{\rm{D}}/T} \frac{x^4 e^x}{(e^x-1)^2}\,dx}
\label{eq:Debye_HC}
\end{equation}
\ese
is the Debye lattice heat capacity at constant volume due to acoustic phonons.  The least-squares fits of the $C_{\rm p}(T)$ data for $1.8~{\rm K} \leq T\leq 300$~K by Eqs.~(\ref{Eqs:AllTCpFit}) are shown by the red curves in Figs.~\ref{fig:HC_BaCo2As2}(a) and~\ref{fig:HC_BaCo2As2}(b) where we used the analytic Pad\'e approximant fitting function \cite{Ryan2012} for the Debye lattice heat capacity integral in Eq.~(\ref{eq:Debye_HC}) and the above values of $\gamma$ that were kept fixed, with  $\Theta_{\rm D}$ the only adjustable parameter. The values $\Theta_{\rm D} = 301(3)$~K and 311(4)~K were found for CoAs-flux and Sn-flux grown crystals, respectively. Even though these two $\Theta_{\rm D}$ values are nearly the same, the systematic deviations of the fits in Fig.~\ref{fig:HC_BaCo2As2}(b) from the data for the Sn-grown crystal are larger than those in Fig.~\ref{fig:HC_BaCo2As2}(a) for the CoAs-grown crystal, suggesting a significant difference between the phonon densities of states versus energy for crystals grown using the two fluxes.

The parameters obtained from the heat capacity fits for ${\rm BaCo_2As_2}$ are summarized in Table~\ref{tab:tableHC} together with those of the other compounds studied below. The Debye temperatures $\Theta_{\rm R}$ obtained from fitting $\rho(T)$ data by the Bloch-Gr\"uneisen model are also listed in Table~\ref{tab:tableHC}  for comparison with the respective $\Theta_{\rm D}$ values, where the $\Theta_{\rm R}$ values are seen to be  roughly a factor of two smaller than the respective $\Theta_{\rm D}$ values.  These large differences suggest that the Bloch-Gr\"uneisen theory for the resistivity arising from electron scattering by acoustic phonons is not appropriate for modeling the high-$T$ $\rho(T)$ data for these compounds. 

\subsection{\label{Sec:BaCo2As2_DOS} Electronic Structure Calculations}

\begin{figure}
\includegraphics[width=3in]{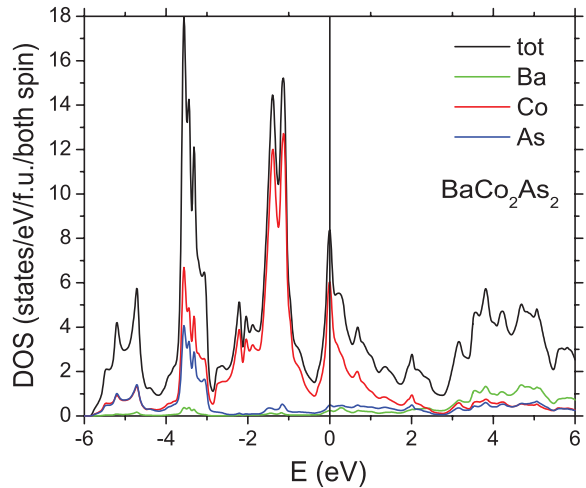}
\caption{(Color online) The density of states (DOS) for ${\rm BaCo_2As_2}$ as a function of energy $E$ relative to the Fermi energy $E_{\rm F}=0$ calculated using FP-LAPW\@. The atom-decomposed DOS of Ba, Co and As atoms are also shown.}
\label{fig:DOS}
\end{figure}

The electronic density of states (DOS) for ${\rm BaCo_2As_2}$ as a function of energy $E$ relative to the Fermi energy $E_{\rm F}$ is shown in Fig.~\ref{fig:DOS}. A similar DOS versus energy has been previously reported for ${\rm BaCo_2As_2}$ by Sefat et al. \cite{Sefat2009} The major contribution to the DOS at $E_{\rm F}$ comes from the Co $3d$ band that is reflected as a sharp peak at $E_{\rm F}$, which was also present in Fig.~4 of Ref.~\onlinecite{Sefat2009}.  The sharp peak in the DOS at $E_{\rm F}$ is due to a Co $d_{x^2-y^2}$ flat band as in ${\rm SrCo_2As_2}$,\cite{Pandey2013b} where the $x$ and~$y$ axes are defined with respect to the Co square lattice. The calculated DOS at $E_{\rm F}$ is
\be
{\cal D}_{\rm band}(E_{\rm F}) = 8.23~{\rm \frac{states}{eV\,f.u.}}
\ee
for both spin directions, which is comparable to but somewhat smaller than the value ${\cal D}_{\rm band}(E_{\rm F}) = 8.5~{\rm states/(eV\,f.u.)}$ found from LDA band calculations in Ref.~\onlinecite{Sefat2009}.  The atom-decomposed contributions to ${\cal D}_{\rm band}(E_{\rm F})$  are Ba: 0.24, Co: 5.92, and As: 0.48 states/(eV~f.u.) for both spin directions. As usual due to the interstitial DOS which is also accounted for in calculation of the total DOS, the total DOS is larger than the sum of the atom-decomposed partial contributions to ${\cal D}_{\rm band}(E_{\rm F})$.

The bare density of states obtained from the DOS calculation is smaller than the value ${\cal D}(E_{\rm F}) = 18.82(2)$~states/eV\,f.u.\ for both spin directions in Table~\ref{tab:tableHC} obtained from the $\gamma$ value for the CoAs-grown crystal, evidently due to the presence of many-body electron-phonon and electron--electron ($m^\ast/m_{\rm band}$) enhancement effects.  We obtain the combined enhancement as
\be
(1+\lambda_{\rm el-ph})\frac{m^\ast}{m_{\rm band}} = \frac{{\cal D}^\gamma(E_{\rm F})}{{\cal D}_{\rm band}(E_{\rm F})} = 2.29,
\label{Eq:BaCo2As2enhance}
\ee
where $\lambda_{\rm el-ph}$ is the electron-phonon coupling constant.  This value is close to the enhancement value of 2.2 obtained for ${\rm BaCo_2As_2}$ in Ref.~\onlinecite{Sefat2009}.  Xu et al.\ found that their ARPES band dispersion data for ${\rm BaCo_2As_2}$ could be reproduced by band theory with only a modest renormalization $m^\ast/m_{\rm band} \approx 1.4$,\cite{Xu2013} which together with Eq.~(\ref{Eq:BaCo2As2enhance}) yields the value $\lambda_{\rm el-ph} \approx 0.6$.  This electron-phonon coupling constant is similar to the value $\lambda_{\rm el-ph} = 0.76$ obtained from {\it ab initio} calculations for the isostructural and closely-related compound ${\rm BaNi_2As_2}$.\cite{Subedi2008}

\subsection{\label{Sec:BaCo2As2_MT-MH} Magnetization and Magnetic Susceptibility}

\begin{figure}
\includegraphics[width=3in]{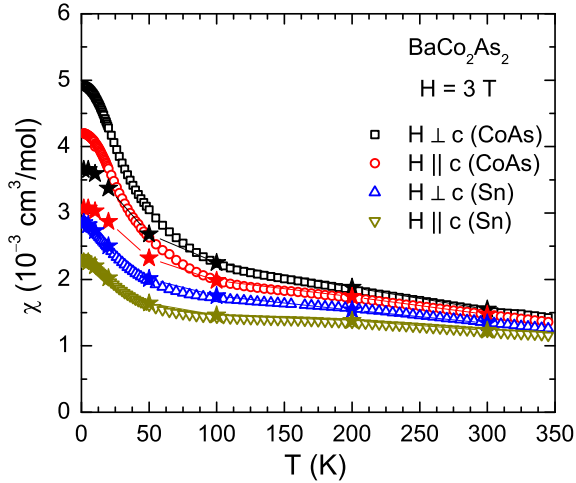}\vspace{0.1in}
\caption{(Color online) Zero-field-cooled magnetic susceptibility $\chi\equiv M/H$ of CoAs~flux-grown and Sn~flux-grown ${\rm BaCo_2As_2}$ single crystals versus temperature $T$ in the $T$ range 1.8--350~K measured in a magnetic field $H = 3.0$~T applied along the $c$-axis ($\chi_c, H \parallel c$) and in the $ab$-plane ($\chi_{ab}, H \perp  c$). The stars of respective colors represent the  $\chi$ obtained from fitting $M(H)$ isotherm data by Eq.~(\ref{eq:MH_linear-fit}). The solid lines connecting the stars are guides to the eye.}
\label{fig:MT_BaCo2As2}
\end{figure}

The $T$ dependences of the zero-field-cooled magnetic susceptibility $\chi \equiv M/H$ of CoAs~flux-grown as well as Sn~flux-grown ${\rm BaCo_2As_2}$ single crystals measured in a magnetic field $H= 3.0$~T applied along the $c$~axis ($\chi_c, H \parallel c$) and in the $ab$~plane ($\chi_{ab}, H \perp  c$) are shown in Fig.~\ref{fig:MT_BaCo2As2}. These data indicate that ${\rm BaCo_2As_2}$ remains paramagnetic without any magnetic phase transitions above 1.8~K, as also deduced previously.\cite{Sefat2009}  A weak $T$ dependence is observed with an upturn in $\chi$ below $\sim 50$~K\@. The $\chi$ of the Sn~flux-grown crystal is smaller than that of CoAs~flux-grown crystal. The $\chi$ of both the CoAs~flux-grown and Sn~flux-grown crystals are anisotropic over the entire 1.8--350~K temperature range of measurement with $\chi_{ab} > \chi_{c}$.  This anisotropy with $\chi_{ab} > \chi_{c}$ was reproducible on several batches of crystals.  The same sign of the anisotropy in $\chi$ is commonly observed in undoped and doped FeAs-based compounds.\cite{Johnston2010}

\begin{figure}
\includegraphics[width=3in]{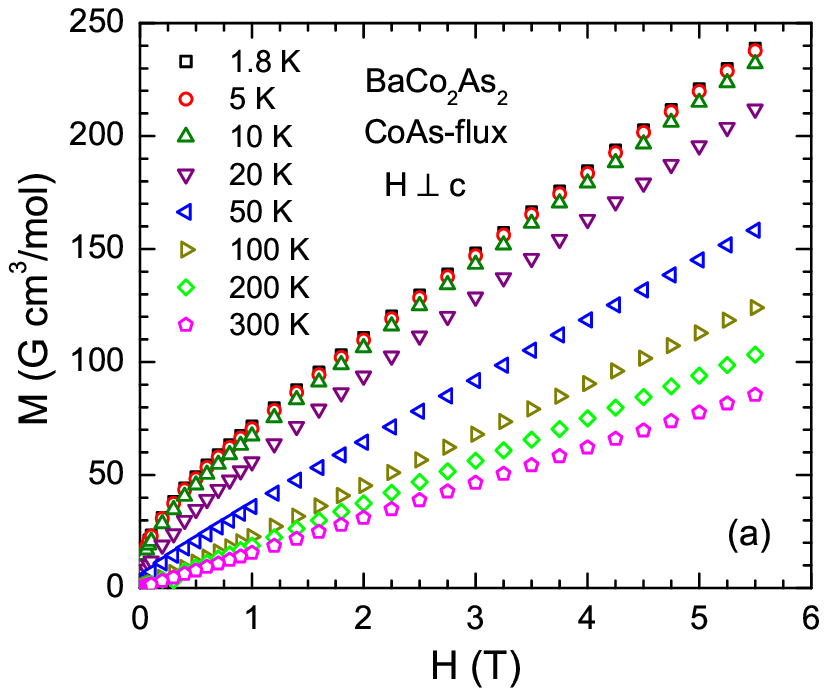}
\includegraphics[width=3in]{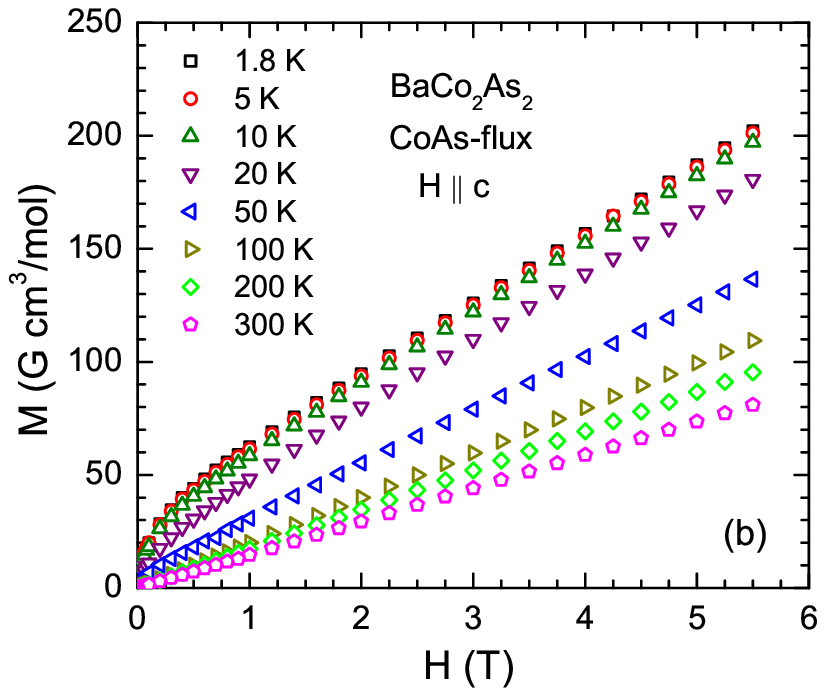}
\caption{(Color online) Isothermal magnetization $M$ of a CoAs~flux-grown ${\rm BaCo_2As_2}$ single crystal as a function of applied magnetic field $H$ measured at the indicated temperatures for $H$ applied (a) in the $ab$~plane ($M_{ab}, H \perp  c$) and (b)~along the $c$~axis ($M_c, H \parallel c$).}
\label{fig:MH_BaCo2As2_CoAs}
\end{figure}

\begin{figure}
\includegraphics[width=3in]{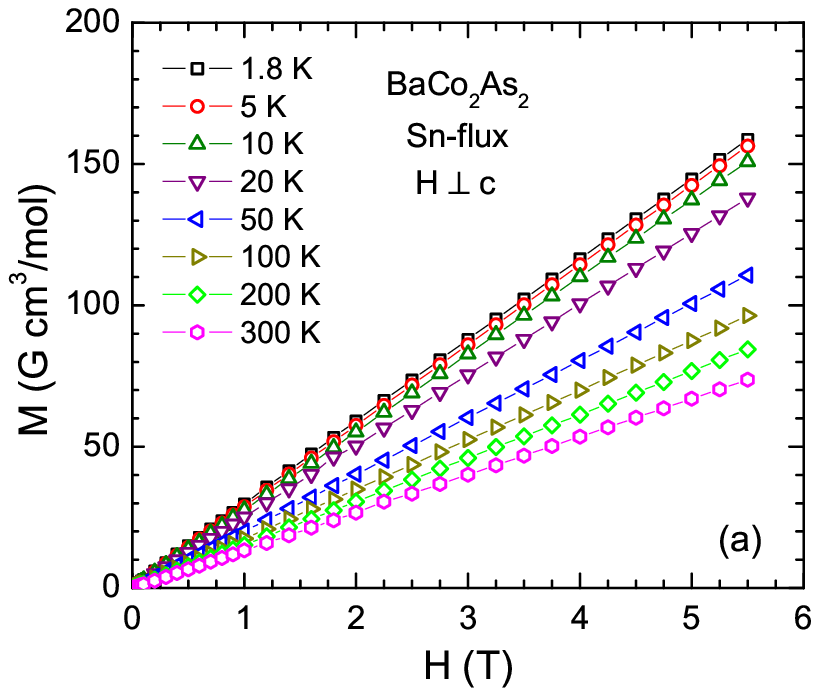}
\includegraphics[width=3in]{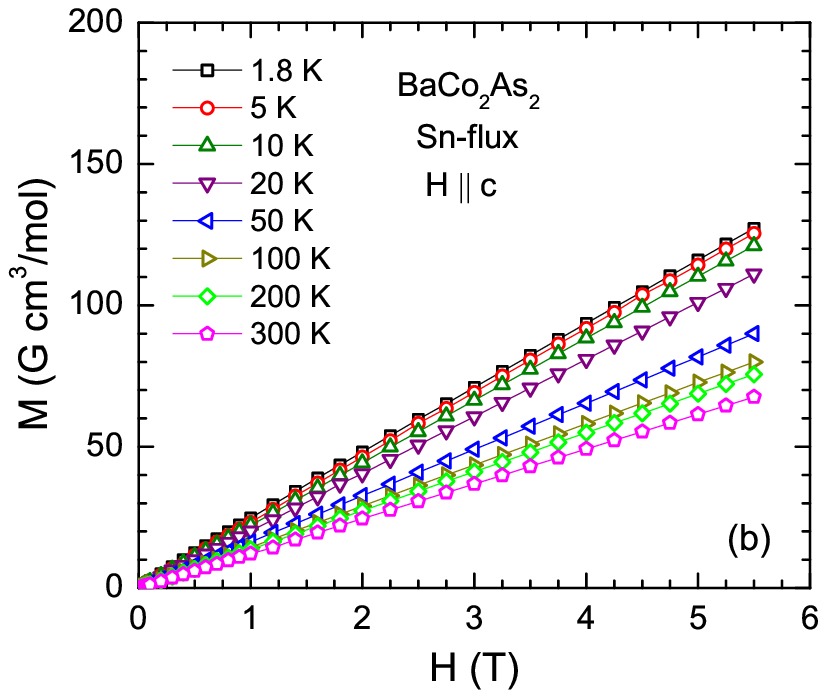}
\caption{(Color online) Isothermal magnetization $M$ of a Sn~flux-grown ${\rm BaCo_2As_2}$ single crystal as a function of applied magnetic field $H$ measured at the indicated temperatures for $H$ applied (a)~in the $ab$~plane ($M_{ab}, H \perp  c$) and (b)~along the $c$~axis ($M_c, H \parallel c$).}
\label{fig:MH_BaCo2As2_Sn}
\end{figure}

The $M(H)$ isotherms obtained for ${\rm BaCo_2As_2}$ at eight temperatures between 1.8 and 300~K for both \mbox{$ H \perp  c$} ($M_{ab}$) and $H\parallel c$ ($M_c$) are shown in Figs.~\ref{fig:MH_BaCo2As2_CoAs} and~\ref{fig:MH_BaCo2As2_Sn} for CoAs~flux-grown and Sn~flux-grown crystals, respectively.  While the $M(H)$ isotherms for the Sn~flux-grown crystal are almost linear in $H$ even at the lowest measured $T = 1.8$~K, the $M(H)$ isotherms of CoAs~flux-grown crystals exhibit nonlinearity at low $H$ at $T \lesssim 20$~K, although the $M(H)$ isotherms for $T \geq 100$~K are almost linear in $H$\@.    The nonlinearity in $M(H)$ was reproducibly found in another batch of ${\rm BaCo_2As_2}$ crystals grown out of CoAs self-flux.

This nonlinearity in the $M(H)$ isotherms in Fig.~\ref{fig:MH_BaCo2As2_CoAs} of CoAs~flux-grown crystals at $T\lesssim 20$~K might be considered to be due to the presence of a small amount of FM impurity. However, the FM impurity cannot be CoAs from the flux, because CoAs is a Pauli paramagnetic metal;\cite{Saparov2012} another potential impurity is CoAs$_2$ but this compound is a diamagnetic semiconductor.\cite{Siegrist1986}  Furthermore, the nonlinearity in the $M(H)$ curves for $H<1$~T does not rise as fast with increasing field as expected from a ferromagnetic impurity which has a spontaneous ordered moment at $H=0$ (see Figs.~\ref{fig:MH_BaKCo2As2}, \ref{fig:MH_BaKCo2As2_lowH} and~\ref{fig:MH_BaKCo2As2_CoAs} below).  A more likely scenario is that the low-field nonlinearity in the $M(H)$ isotherms of the CoAs~flux-grown crystals at $T \lesssim 20$~K is due to the presence of saturable paramagnetic defects.  In this case the absence of such nonlinearities in the $M(H)$ data for the Sn~flux-grown crystals would be associated with absence of paramagnetic defects in these crystals.  This inference is consistent with the different crystal growth temperature profiles for the FeAs- and Sn-grown crystals because the FeAs-grown crystals were quenched from a high temperature of 1180~$^\circ$C, whereas the Sn-grown crystals were quenched from the much lower temperature of 600~$^\circ$C, as discussed in Sec.~\ref{ExpDetails}.  The lower quenching temperature would tend to anneal out paramagnetic defects present at higher temperatures.

The intrinsic susceptibility $\chi$ of a CoAs~flux-grown ${\rm BaCo_2As_2}$ crystal is the slope of the high-field linear part of am $M(H)$ isotherm where the paramagnetic defects are saturated, and is obtained from a fit of the $M(H)$ data for $H \geq 2$~T by
\begin{equation}
M(H) = M_{\rm s} + \chi H,
\label{eq:MH_linear-fit}
\end{equation}
where $M_{\rm s}$ is the paramagnetic defect saturation magnetization.  The resulting $\chi$ data versus temperature are shown as stars in Fig.~\ref{fig:MT_BaCo2As2}. 

For the CoAs~flux-grown crystal the intrinsic $\chi$ values obtained from the slopes of the high-field fits of the $M(H)$ data in Fig.~\ref{fig:MH_BaCo2As2_CoAs} by Eq.~(\ref{eq:MH_linear-fit}) decrease slowly with increasing $T$ for $T\geq 100$~K with negligible saturation magnetization values $M_{\rm s}^{ab} = 0.0(5)$~G\,cm$^3$/mol and $M_{\rm s}^{c} = 0.0(4)$~G\,cm$^3$/mol\@.  For this reason these $\chi(T)$ values, shown as the stars in Fig.~\ref{fig:MT_BaCo2As2}, agree with the values of $M(T)/H$ data in Fig.~\ref{fig:MT_BaCo2As2} in this $T$ range.  On the other hand, the intrinsic $\chi$ values for $T\lesssim 20$~K are significantly smaller than obtained from the $M(T)/H$ measurements as seen in Fig.~\ref{fig:MT_BaCo2As2}, a difference that arises from the low-field nonlinearity of the $M(H)$ isotherms in Fig.~\ref{fig:MH_BaCo2As2_CoAs} in this temperature range.  We concude that intrinsic upturns in the $\chi(T)$ data occur below $\sim 100$~K in Fig.~\ref{fig:MT_BaCo2As2}.  This conclusion is confirmed by the NMR shift measurements of the $T$-dependent spin susceptibility of a CoAs~self-flux-grown crystal in the following section.

By fitting the $M(H)$ data for the CoAs~flux-grown crystal at $T=1.8$~K and $H>2$~T in Fig.~\ref{fig:MH_BaCo2As2_CoAs} by Eq.~(\ref{eq:MH_linear-fit}), one obtains a saturation magnetization of the paramagnetic defects for either field direction as $M_{\rm s} \approx 40~{\rm G\,cm^3/mol} \approx 0.0072~\mu_{\rm B}$/f.u., which corresponds to the saturation magnetization of 0.72~mol\% of spin-1/2 defects with spectroscopic splitting factor $g=2$.

For the Sn~flux-grown crystals, good agreement is seen in Fig.~\ref{fig:MT_BaCo2As2} between the $\chi\equiv M(T)/H$ data and the $\chi$ values obtained from the high-field slope of the $M(H)$ isotherm data in Fig.~\ref{fig:MH_BaCo2As2_Sn}  for the entire 1.8 to 300~K temperature range, indicating that the $M_{\rm s}$ values are negligible.  Thus there are no observable FM impurities or paramagnetic defects in Sn~flux-grown BaCo$_2$As$_2$ crystals.

\subsection{NMR Spectroscopy}

\subsubsection{$^{75}$As NMR Spectra}

\begin{figure}[t]
\includegraphics[width=3.3in]{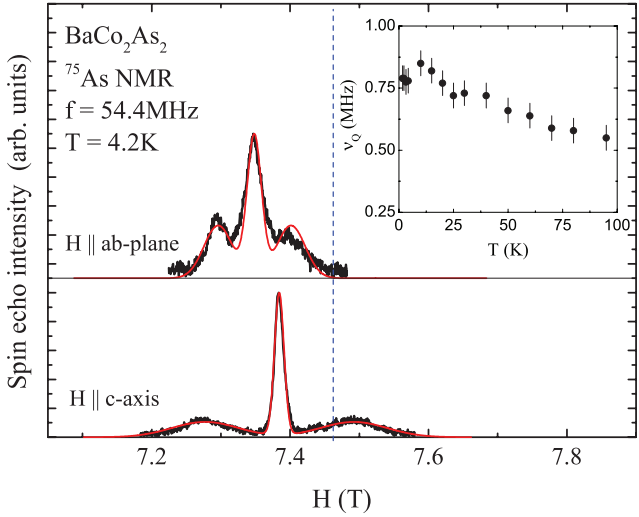} 
\caption{(Color online) Field-swept $^{75}$As-NMR spectra at temperature $T$ = 4.2 K for magnetic fields perpendicular (top panel) and parallel (bottom panel) to the $c$~axis.  The black and red lines are observed and simulated spectra, respectively.  The vertical dotted line corresponds to the zero-shift ($K = 0$) position. Inset: $^{75}$As nuclear quadruple frequency $\nu_{\rm{Q}}$ versus~$T$.}
\label{Fig:BaCo2As2NMRspectra}
\end{figure}

Figure~\ref{Fig:BaCo2As2NMRspectra} shows field-swept $^{75}$As-NMR spectra of a CoAs~flux-grown BaCo$_2$As$_2$ crystal at $T = 4.2$~K for magnetic fields $H \parallel c$~axis and $H\parallel ab$~plane.  The spectra exhibit features typical of a nuclear spin $I = 3/2$ with Zeeman and quadrupolar interactions, which result in a sharp central transition and two satellite peaks split by the quadrupolar interaction of the As nucleus with the local electric field gradient (EFG).  The observed quadrupole-split NMR spectra were reproduced by the nuclear spin Hamiltonian 
\be
{\cal H} = -\gamma\hbar\vec{I}\cdot\vec{H}_{\rm eff} + \frac{h \nu_{\rm Q}}{6} [3I_{\rm z}^{2}-I(I+1)],
\label{Eq:IHam}
\ee
where $H_{\rm eff}$ is the effective field at the As site (summation of external field $H$ and 
the hyperfine field $H_{\rm hf}$), $h$ is Planck's constant, $\hbar = h/(2\pi)$ and $\nu_{\rm Q}$ is the nuclear quadrupole frequency.  Here the asymmetry parameter of the EFG is assumed to be zero because of the tetragonal symmetry of the compound.  The red curves in Fig.~\ref{Fig:BaCo2As2NMRspectra} show simulated spectra calculated from Hamiltonian~(\ref{Eq:IHam}).  The principle axis of the EFG at the As site in tetragonal BaCo$_2$As$_2$ is along the crystal $c$~axis as in SrCo$_2$As$_2$ (Ref.~\onlinecite{Pandey2013b}) and also as in members of the $A$Fe$_2$As$_2$ family with tetragonal symmetry\cite{Kitagawa2009, Kitagawa2008, Baek2009, Ning2009, Urbano2010} since the As site has a local fourfold rotational symmetry about the $c$~axis.  The satellite linewidth, which reflects the distribution of EFG due to defects or lattice distortion, is relatively large.  To reproduce the linewidth, one needs to introduce $\sim 63 $\% distributions of the quadrupole frequency ($\Delta \nu_{\rm Q}$ = 0.5 MHz)  at $T = 4.2$~K as shown by red curves in the figure.  This is much larger than the $\sim 7 $\% distributions of $\nu_{\rm Q}$ = 8.45 MHz at $T = 4.2$~K for the case of $^{75}$As-NMR in SrCo$_2$As$_2$.\cite{Pandey2013b}  This indicates  that the local As environment in BaCo$_2$As$_2$ has a lower degree of homogeneity than in SrCo$_2$As$_2$.  The spectra for BaCo$_2$As$_2$ for both $H$ directions did not show any obvious magnetic broadenings at any temperature between 1.6 and 275~K, which indicates that static magnetic ordering does not occur in this compound above 1.6~K\@. 

The inset of the Fig.~\ref{Fig:BaCo2As2NMRspectra} shows the $T$ dependence of the nuclear quadruple frequency $\nu_{\rm{Q}}$ =  $eQV_{ZZ}$/2$h$ where $e$ is the fundamental charge, $Q$ is the quadrupole moment of the As nucleus and $V_{ZZ}$ is the EFG at the As site.  The  $\nu_{\rm{Q}}$  increases by about 45\% from 0.55~MHz at 95~K to 0.8~MHz at 1.6~K\@.  This behavior contrasts sharply with that observed in SrCo$_2$As$_2$  where $\nu_{\rm{Q}}$ decreases from 10~MHz at 250~K to 8.45~MHz at 1.6~K\@. An increase of  $\nu_{\rm{Q}}$ by $\sim$ 30\% on lowering $T$ from 2~MHz at 300~K to 2.6~MHz at $\sim$ 210~K was observed in SrFe$_2$As$_2$.\cite{Kitagawa2009}  Since $V_{ZZ}$ arises from hybridization between the As-4$p$ and Co-3$d$ orbitals with an additional contribution from the noncubic part of the spatial distribution of surrounding ions, the small $\nu_{\rm{Q}}$ in BaCo$_2$As$_2$ compared with that in SrCo$_2$As$_2$ indicates a weaker hybridization between the orbitals in BaCo$_2$As$_2$.  This can be qualitatively interpreted when one takes into account the difference of $c$-axis lattice constant of 12.67~\AA\ in BaCo$_2$As$_2$ and 11.80~\AA\ in SrCo$_2$As$_2$.

\begin{figure}[t]
\includegraphics[width=8cm]{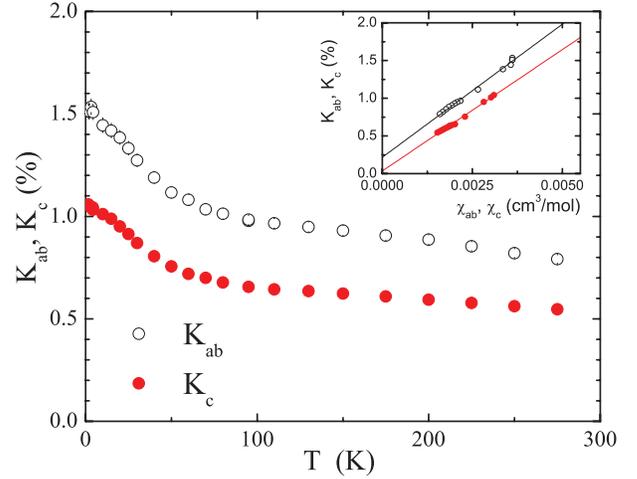} 
\caption{(Color online) (a) $T$-dependence of $^{75}$As NMR shifts $K_{\rm ab}$ and $K_{\rm c}$ for BaCo$_2$As$_2$.  Inset: $K$ versus magnetic susceptibility $\chi$ plots for the corresponding $ab$ and $c$ components of $K$ and $\chi$ with $T$ as an implicit parameter, where we used $\chi$ data shown by stars in Fig.~\ref{fig:MT_BaCo2As2} below 150~K and the observed $\chi$ for $T\geq 150$~K\@.  The solid lines are linear fits for the two field directions. }
\label{Fig:BaCo2As2NMR_K}
\end{figure}

Figure~\ref{Fig:BaCo2As2NMR_K} shows the $T$ dependence of the NMR shifts $K_{ab}$ and $K_{c}$ for $H$  
parallel to the $ab$ plane and to the $c$~axis, respectively.  We emphasize that these results are intrinsic to the crystal and are not sensitive to the presence of the paramagnetic defects in CoAs-grown crystals detected in the above $\chi(T)$ measurements.  With decreasing $T$, both $K_{ab}$ and $K_{c}$ increase slowly down to $\sim  50$~K and show a significant additional increase below that temperature, similar to the corresponding intrinsic $\chi_{ab}$ and $\chi_c$ data for a CoAs~flux-grown crystal of ${\rm BaCo_2As_2}$ (the stars in Fig.~\ref{fig:MT_BaCo2As2}).  The NMR shift consists of the $T$-dependent spin shift $K_{\rm spin}(T)$ and the $T$-independent orbital (chemical) shift $K_{\rm chem}$: $K(T) =K_{\rm spin}(T) + K_{\rm chem}$, where $K_{\rm spin}(T)$ is proportional to the intrinsic spin susceptibility ${\chi_{\rm spin}(T)}$ via the hyperfine coupling constant $A$, $K_{\rm spin}(T)$  = $\frac{A\chi_{\rm spin}(T)}{N_{\rm A}}$.  Here  ${N_{\rm A}}$ is Avogadro's number.  As shown in the inset of Fig.~\ref{Fig:BaCo2As2NMR_K}, both $K_{ab}$ and $K_{c}$ indeed vary linearly with the corresponding intrinsic $\chi$ where we used $\chi$ data shown by stars in Fig.~\ref{fig:MT_BaCo2As2} below 150~K and the observed $\chi$ for $T \geq 150$~K\@.  From the respective slopes, the hyperfine coupling constants are estimated to be $A_{ab} = (40 \pm 3)$~kOe/$\mu_{\rm B}$  and $A_{c} = (36 \pm 2)$~kOe/$\mu_{\rm B}$, respectively.  These values are respectively smaller than $A_{ab} = 65.9$~kOe/$\mu_{\rm B}$  and $A_{c} = 45.0$~kOe/$\mu_{\rm B}$ for SrCo$_2$As$_2$.\cite{Pandey2013b}  Since the hyperfine coupling is mainly due to hybridization between the $4s$ and $4p$ orbitals of the As atoms with the $3d$ orbitals of the Co atoms, the small hyperfine couplings indicate a weak hybridization between these orbitals in BaCo$_2$As$_2$.  This is consistent with the small $\nu_{\rm{Q}}$ for $^{75}$As in BaCo$_2$As$_2$ because the $\nu_{\rm{Q}}$ is also affected by the strength of the hybridization between the As-$4p$ and Co-$3d$ orbitals.  As pointed out above, a weak hybridization between the orbitals can be qualitatively explained by taking into account the difference between the $c$~axis lattice constants in BaCo$_2$As$_2$ and SrCo$_2$As$_2$. 

The agreement between the respective $T$ dependences for both field directions of the intrinsic $\chi(T)$ in Fig.~\ref{fig:MT_BaCo2As2} of a CoAs-grown BaCo$_2$As$_2$ crystal with those of the NMR $K(T)$ of a similar crystal demonstrates that the former $T$ dependences are the intrinsic $T$-dependences of the spin susceptibilities of BaCo$_2$As$_2$.  These $T$ dependences are  different from those of pure and doped BaFe$_2$As$_2$ crystals above their long-range AFM spin-density wave ordering temperatures,\cite{Johnston2010} where the $\chi(T)$ data for both field directions decrease with decreasing~$T$ instead of increasing with decreasing~$T$ as in BaCo$_2$As$_2$.

$K_{\rm chem}$ is estimated to be $(0.22 \pm 0.02)$\% and $(0.04 \pm 0.02)$\% for $H$ parallel to the $ab$ plane and $c$~axis, respectively.  The $K_{{\rm chem},c}$  is similar to $K_{{\rm chem},c}\sim 0$ in SrCo$_2$As$_2$, and  
$K_{{\rm chem,} ab}$ is larger than $K_{{\rm chem,} ab}\sim 0$  in SrCo$_2$As$_2$.  On the other hand,  $K_{{\rm chem,} ab} = 0.22$\% is  close to $K_{\rm chem}\sim 0.2 $\% reported in BaFe$_2$As$_2$ (Ref.~\onlinecite{Kitagawa2008}) and $K_{\rm chem}\sim 0.22$\% in BaFe$_{1.84}$Co$_{0.16}$As$_2$.\cite{Ning2010}  Kitagawa et~al.\ commented that the value of $K_{\rm chem}\sim0.2$\% seemed rather large.\cite{Kitagawa2008}  At present the origin of the compound- and orientation-dependent $K_{\rm chem}$ is not clear. 

\subsubsection{$^{75}$As Nuclear Spin Dynamics}

\begin{figure}[t]
\includegraphics[width=3.3in]{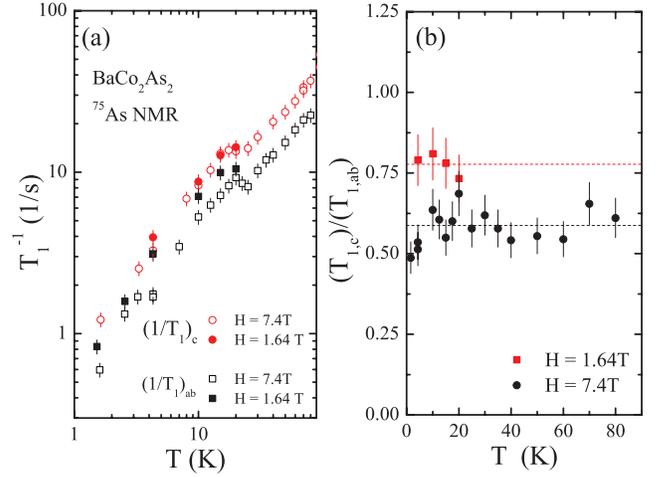} 
\caption{(Color online) (a) $^{75}$As nuclear spin-lattice relaxation rate 1/$T_1$ versus temperature~$T$ for BaCo$_2$As$_2$ for magnetic field directions $H$ $\parallel$ $c$~axis and $H$ $\parallel$ $ab$~plane under two magnetic fields $H = 1.64$ and 7.4~T\@.  (b) $T$-dependences of the ratio $r\equiv T_{\rm 1,c}/T_{\rm 1, ab}$  for $H = 1.64$ and 7.4~T\@.  }
\label{Fig:BaCo2As2NMR_T1}
\end{figure}

The $^{75}$As nuclear spin-lattice relaxation rate 1/$T_1$ at each $T$ is determined by fitting the nuclear magnetization $M$ versus time $t$ dependence of the central line after saturation using the double-exponential function 
\be
1-\frac{M(t)}{M({\infty})} = 0.1\exp(-t/T_{\rm 1}) + 0.9\exp(-6t/T_{\rm 1}) ,
\ee
as expected for the central line of the spectrum of the  $^{75}$As ($I$ = 3/2) nucleus, where $M(t)$ and $M({\infty})$  are the nuclear magnetization at time $t$ after saturation and the equilibrium nuclear magnetization at time $t\rightarrow\infty$, respectively. Figure~\ref{Fig:BaCo2As2NMR_T1}(a) shows 1/$T_1$ versus $T$ for magnetic fields parallel and perpendicular to the $c$~axis at $H = 7.4$~T\@.   For both field orientations, the 1/$T_{\rm 1}$ values decrease monotonically with decreasing $T$, show slight increases around 20~K and then decrease again at low temperatures.  $1/T_1$ was also measured at a low magnetic field $H = 1.64$~T as shown in Fig.~\ref{Fig:BaCo2As2NMR_T1}(a), where the 1/$T_1$ data are limited to temperatures below $\sim 20$ K due to poor signal intensity at higher temperatures at the low magnetic field.  As can be seen in the figure,  1/$T_{1c}$ is independent of $H$ within our experimental uncertainty, but 1/$T_{1,ab}$ increases with decreasing $H$\@.

Based on these results, we now discuss the relationship between the anisotropy of our 1/$T_1$ data and the anisotropy of the Co spin fluctuations.  We closely follow the procedure previously developed for Fe pnictides, \cite{Kitagawa2009, Kitagawa2010, Fukazawa2012} and especially for the recent $^{75}$As NMR measurements of SrCo$_2$As$_2$.\cite{Pandey2013b}  Defining the amplitude of the spin fluctuation spectral density as ${\cal S}_{\alpha}({\bf q}, \omega_{\rm N})$ ($\alpha$ = $ab$, $c$), where {\bf q} is the wavevector of a spin fluctuation in terms of Co square lattice notation,  one obtains\cite{Pandey2013b,Kitagawa2010}
\bea
\begin{array}{cc}
1/T_{1,ab}  \\
1/T_{1,c}  
\end{array}
=
\begin{cases}
\begin {array}{cc}
|A_{ab}{\cal S}_{ab}|^2 + |A_c{\cal S}_{c}|^2\\
2|A_{ab}{\cal S}_{ab}|^2
\end {array}
\hspace{0.2in}[{\bf q} = (0,0)]\vspace{0.1in}\\
\begin {array}{cc}
|A_{ab}{\cal S}_{ab}|^2 + |A_c{\cal S}_{c}|^2\\
2|A_{ab}{\cal S}_{ab}|^2
\end {array}
 [{\bf q} = (\pi,0),\ (0,\pi)]\vspace{0.1in}\\
\begin {array}{ccc}
|D{\cal S}_{ab}|^2  \\
2|D{\cal S}_{ab}|^2  \\
\end {array}
\hspace{0.83in} [{\bf q} = (\pi,\pi)],
\label{Eqs:1T1}
\end{cases}
\eea
where {\bf q} = (0,0) is the $ab$-plane component of the ferromagnetic wavevector, ${\bf q} = (\pi,0)$ or $(0,\pi)$ is the stripe wave vector as observed in the $A{\rm Fe_2As_2}$ and ${\rm SrCo_2As_2}$ compounds and ${\bf q} = (\pi,\pi)$ is the alternately-termed N\'eel-, checkerboard- or G-type AFM wavevector.  Then defining the ratio
\be
r \equiv \frac{1/T_{1,ab}}{1/T_{1,c}}, 
\ee
Eqs.~(\ref{Eqs:1T1}) give
\begin{eqnarray}
r =  \left\{
\begin {array}{lll}
0.5 + 0.5 \left(\frac{A_c {\cal S}_c}{A_{ab} {\cal S}_{ab}}\right) ^2 \hspace{0.1in} [{\bf q} = (0,0)] \vspace{0.1in} \\
0.5 + \left(\frac{{\cal S}_{ab}}{{\cal S}_c}\right) ^2  \hspace{0.3in}  [{\bf q} = (\pi,0),\ (0, \pi)] \vspace{0.1in}\\ 
0.5  \hspace{1.1in}   [{\bf q} = (\pi,\pi)]. \\
\end {array}
\right .
\label{eqn:correlation}
\end{eqnarray}

As plotted in Fig.~\ref{Fig:BaCo2As2NMR_T1}(b), the $r$ value is almost constant versus $T$ with the magnetic field dependent values of $r$ $\approx$ 0.78 and 0.59 for $H$ = 1.64 and 7.4~T, respectively.  We first analyze the $T_1$ data with the first of Eqs.~(\ref{eqn:correlation}).  Using the ratio of the hyperfine fields $A_{c}/A_{ab}$  = 36/40 = 0.9 determined from the $K$-$\chi$ plots in the inset of Fig.~\ref{Fig:BaCo2As2NMR_K}, the values of $r$ $\approx$ 0.78 and 0.59 provide anisotropic spin fluctuations with ${\cal S}_{c}$ = 0.83 ${\cal S}_{ab}$ and ${\cal S}_{c}$ = 0.47 ${\cal S}_{ab}$ for $H$ = 1.64 and 7.4~T, respectively.  This means that ferromagnetic Co spin fluctuations in the $ab$-plane are stronger than along the $c$-axis.  Anisotropic ferromagnetic spin fluctuations were also pointed out in SrCo$_2$As$_2$,\cite{Pandey2013b} where ${\cal S}_{c}$ is larger than ${\cal S}_{ab}$ in contrast to our result for BaCo$_2$As$_2$.  As for the magnetic field dependence, Eqs.~(\ref{Eqs:1T1}) show that the $H$-independent 1/$T_{1,c}$ along the $c$-axis indicates that ${\cal S}_{ab}$ is not affected by the application of $H$, thus ${\cal S}_{c}$ can be concluded to be suppressed with $H$\@.  N\'eel-type spin fluctuations can be ruled out because according to Eqs.~(\ref{Eqs:1T1}), $r$ must be independent of $H$, in conflict with our measurements which show an $H$-dependent~$r$. 

Our 1/$T_1$ data can also be analyzed with the second of Eqs.~(\ref{eqn:correlation}) in terms of stripe-type spin fluctuations with ${\bf q} = (\pi, 0)$ or $(0, \pi$).  Such spin fluctuations/correlations have been revealed in SrCo$_2$As$_2$ from inelastic neutron scattering measurements.\cite{Jayasekara2013}  Using Eqs.~(\ref{eqn:correlation}), the values of $r\approx 0.78$ and 0.59 can be reproduced with anisotropic stripe-type AFM spin fluctuations with ${\cal S}_{ab} = 0.52 {\cal S}_{c}$ and ${\cal S}_{ab} = 0.28 {\cal S}_{c}$ for  $H = 1.64$ and 7.4~T, respectively.  From Eqs.~(\ref{Eqs:1T1}), ${\cal S}_{c}$ is independent of $H$ from the 1/$T_{1,c}$ data, so we conclude that ${\cal S}_{ab}$ is suppressed by $H$ for the stripe-type AFM spin fluctuations.  An anisotropy in stripe-type spin fluctuations is also observed in various Fe-based superconductors near $T_{\rm N}$ in the paramagnetic state, but ${\cal S}_{ab}$ is always larger than ${\cal S}_c$,\cite{Kitagawa2009, Nakai2012} in contrast to the our results.

Although it is clear that the spin fluctuations are anisotropic and $H$-dependent in BaCo$_2$As$_2$, one cannot uniquely determine the nature of the spin fluctuations from the NMR data alone.  Indeed, spin fluctuations may occur with peaks at both the FM and stripe-type AFM wave vectors in the {\bf q}-dependent static susceptibility $\chi({\bf q})$ obtained from local density approximation calculations such as reported for SrCo$_2$As$_2$ in Ref.~\onlinecite{Jayasekara2013}.

\section{\label{BaKCo2As2} Physical Properties of B\lowercase{a}$_{0.946}$K$_{0.054}$C\lowercase{o}$_2$A\lowercase{s}$_2$}

\begin{figure}
\includegraphics[width=3in]{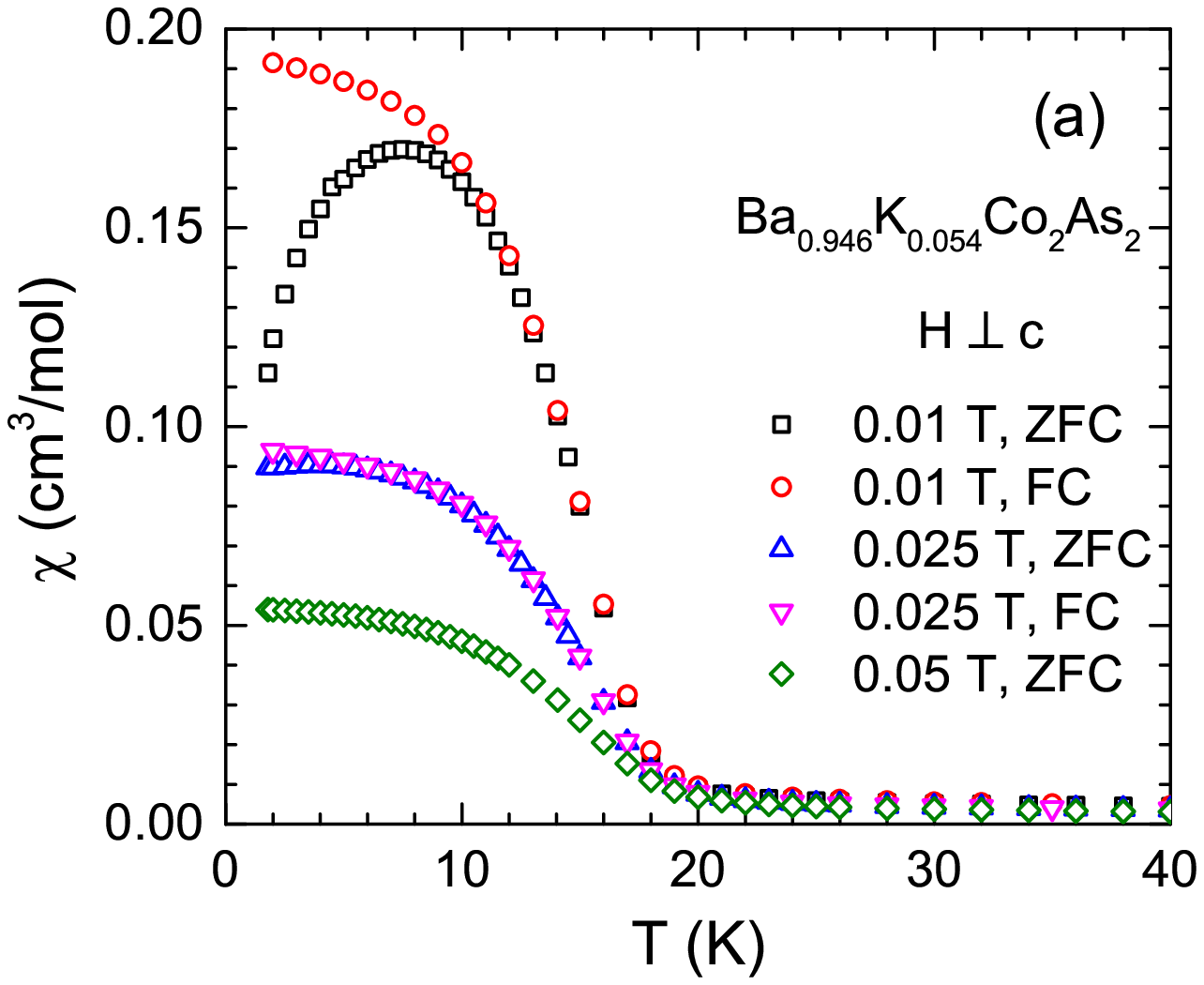}
\includegraphics[width=3in]{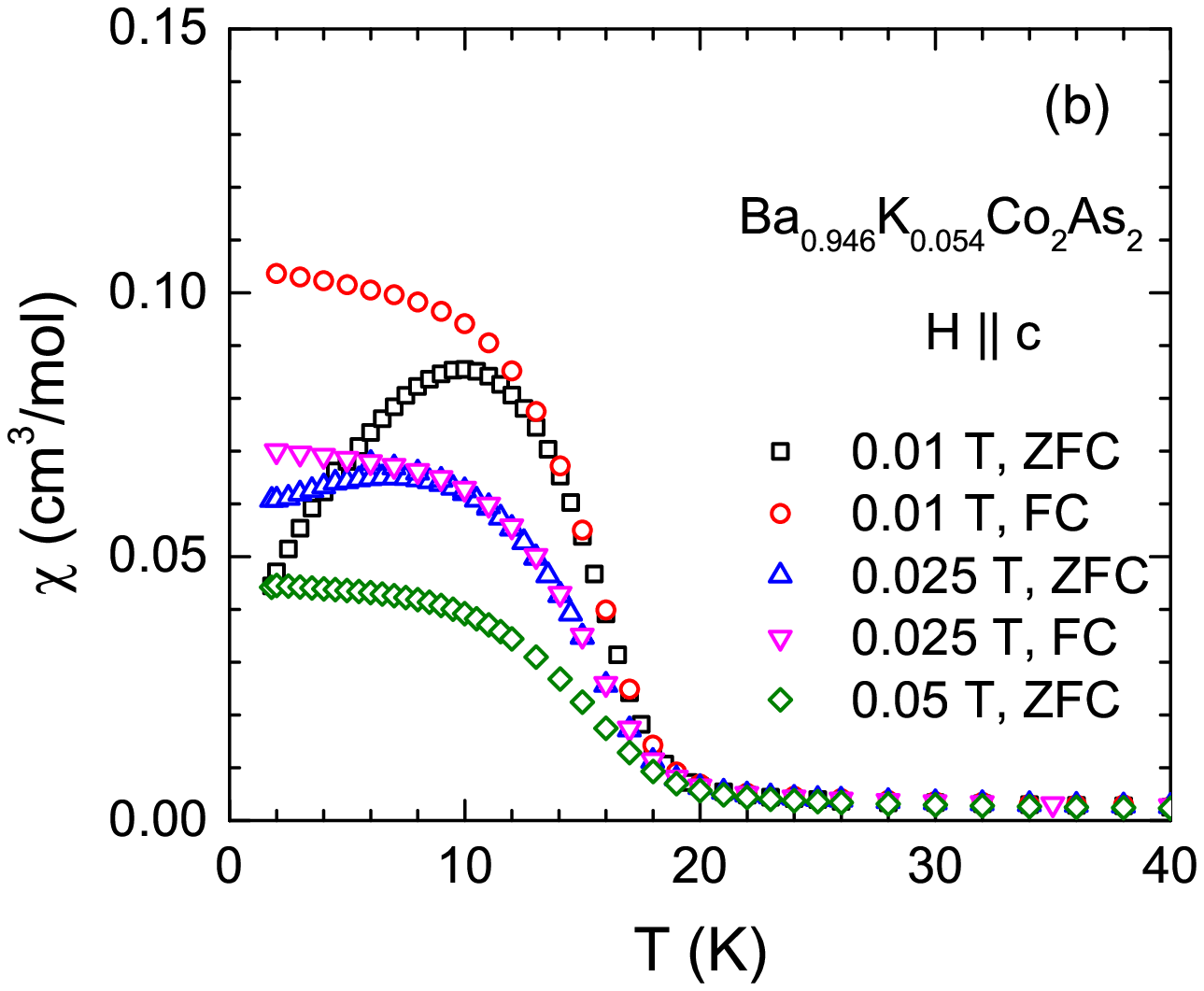}
\caption{(Color online) Zero-field-cooled (ZFC) and field-cooled (FC) magnetic susceptibility $\chi$ of a ${\rm Ba_{0.946}K_{0.054}Co_2As_2}$ single crystal grown in CoAs~flux as a function of temperature $T$ in the temperature range 1.8--40~K measured in low magnetic fields $H$ applied (a) in the $ab$~plane ($\chi_{ab}, H \perp  c$) and, (b) along the $c$~axis ($\chi_c, H \parallel c$).}
\label{fig:MT_BaKCo2As2_Low-H}
\end{figure}

\begin{figure}
\includegraphics[width=3in]{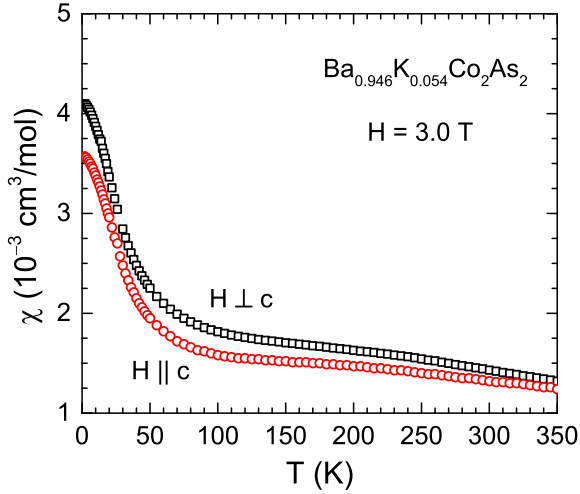}
\caption{(Color online) Zero-field-cooled magnetic susceptibility $\chi$ of a ${\rm Ba_{0.946}K_{0.054}Co_2As_2}$ single crystal grown in CoAs~flux as a function of temperature $T$ in the temperature range 1.8--300~K measured in a magnetic field $H= 3.0$~T applied along the $c$~axis ($\chi_c, H \parallel c$) and in the $ab$~plane ($\chi_{ab}, H \perp  c$).}
\label{fig:MT_BaKCo2As2}
\end{figure}

Single crystals of ${\rm Ba_{0.946}K_{0.054}Co_2As_2}$ were grown in CoAs self-flux.  The $\chi(T) \equiv M(T)/H$ data obtained for a single crystal at low fields $H \leq 0.05$~T are shown in Fig.~\ref{fig:MT_BaKCo2As2_Low-H}. Figure~\ref{fig:MT_BaKCo2As2_Low-H} shows that at $H=0.01$~T a sharp increase in $\chi$ occurs below 18~K with a peak near 10~K in $\chi(T)$ measured on warming after zero-field-cooling (ZFC) for both $H \parallel c$ and $H \perp c$. An irreversibility is observed between the ZFC and field-cooled (FC) $\chi$ data below 12.5~K\@. Such behaviors are typically observed in systems with ferromagnetic (FM) correlations.  An increase in $H$ causes the peak in $\chi(T)$  to broaden and $\chi$ becomes almost $T$ independent below 8~K for $H\geq0.05$~T\@.   No superconductivity is observed at $T \geq1.8$~K\@.  The anisotropy of $\chi(T)$ measured in a much larger field $H=3.0$~T is presented in Fig.~\ref{fig:MT_BaKCo2As2}. An anisotropy $\chi_{ab} > \chi_c$ is observed over the entire $T$~range of the measurement.

\begin{figure} 
\includegraphics[width=3in]{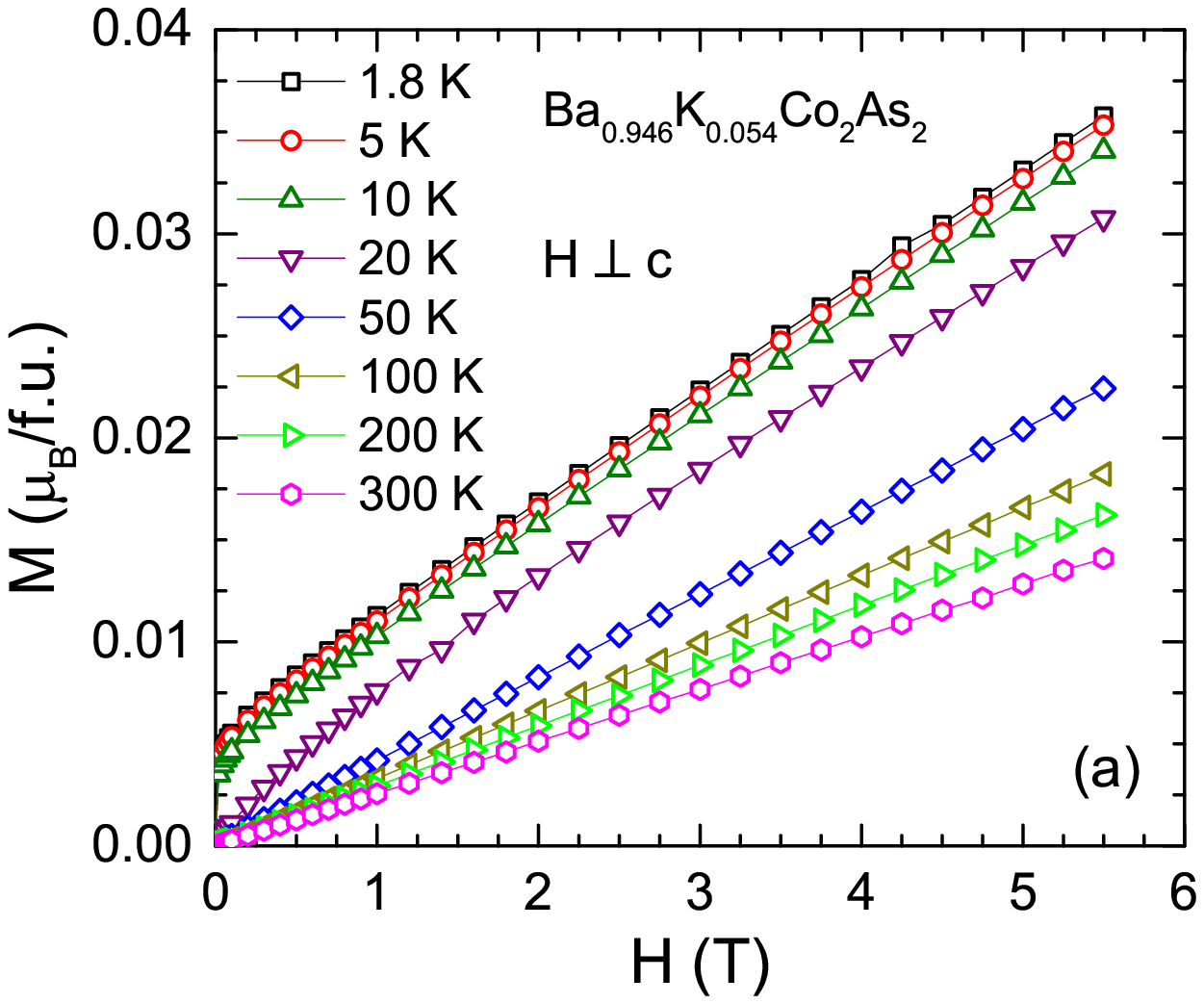}
\includegraphics[width=3in]{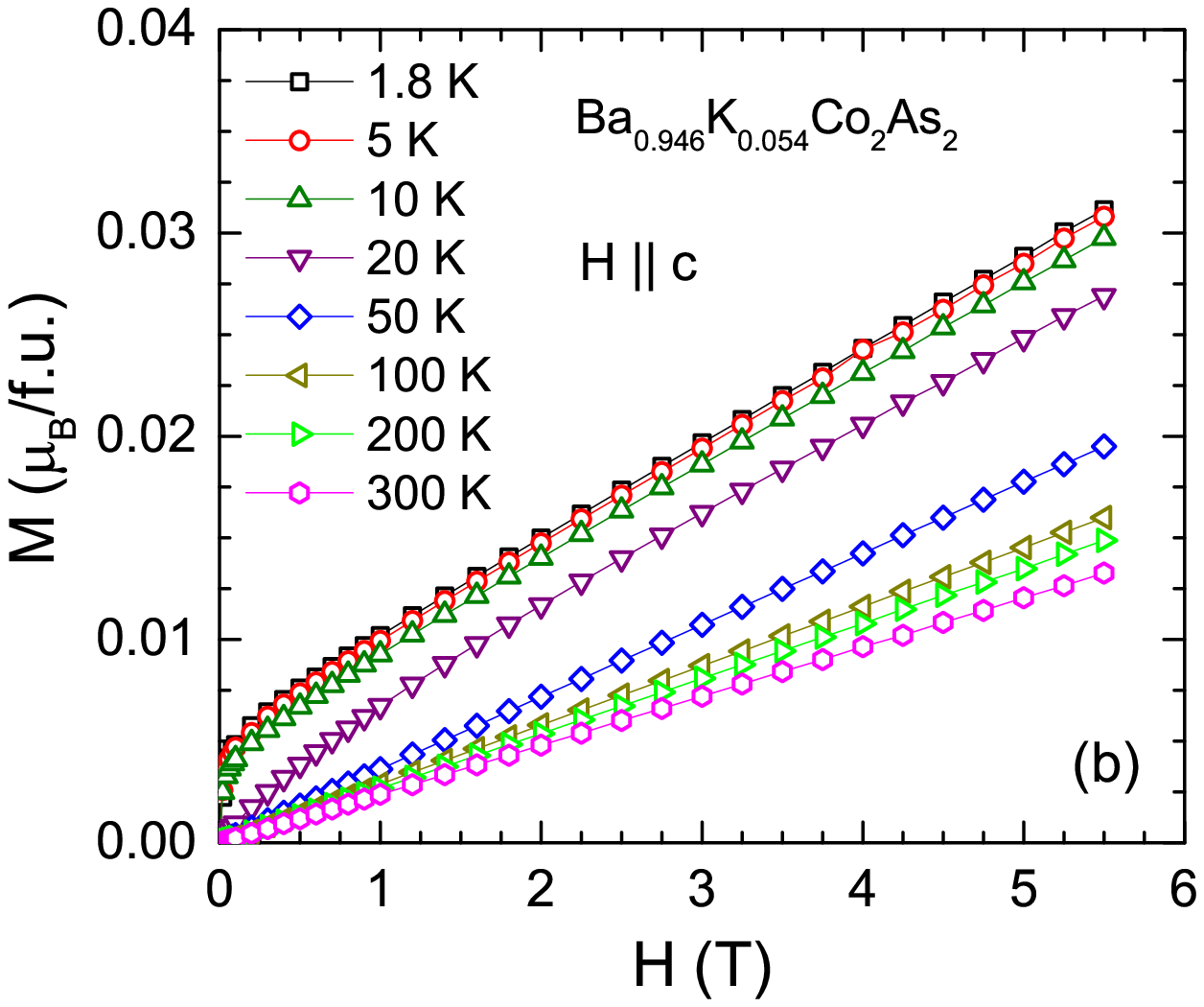}
\caption{(Color online) Isothermal magnetization $M$ of a ${\rm Ba_{0.946}K_{0.054}Co_2As_2}$ single crystal grown in CoAs~flux as a function of applied magnetic field $H$ measured at the indicated temperatures for $H$ applied (a) in the $ab$-plane ($M_{ab}, H \perp  c$) and, (b) along the $c$-axis ($M_c, H \parallel c$).}
\label{fig:MH_BaKCo2As2}
\end{figure}

The $M(H)$ isotherm data for a ${\rm Ba_{0.946}K_{0.054}Co_2As_2}$ single crystal at different temperatures are shown in Fig.~\ref{fig:MH_BaKCo2As2}. For $T>50$~K, the $M(H)$ isotherms are almost linear in $H$ with anisotropy $M_{ab} > M_c$, consistent with the $\chi(T)$ data in Fig.~\ref{fig:MT_BaKCo2As2}.  The appearance of a spontaneous FM moment at low~$H$ is clearly seen in the $M(H)$ isotherms at temperatures $T \leq 10$~K, which are below the temperature of $\sim18$~K at which a FM-like increase is observed in $\chi(T)$ in Fig.~\ref{fig:MT_BaKCo2As2_Low-H}.  The saturation (ordered) moment is obtained by fitting the high-field linear $M(H)$ data at 1.8~K in Fig.~\ref{fig:MH_BaKCo2As2} by Eq.~(\ref{eq:MH_linear-fit}), yielding $M_{\rm s} \approx 0.007~\mu_{\rm B}$/f.u.\ for both $H \perp c$ and $H \parallel c$.  Interestingly, this is the same $M_{\rm s}$ as found above from Fig.~\ref{fig:MH_BaCo2As2_CoAs} for the paramagnetic defects in a CoAs-flux grown crystal of ${\rm BaCo_2As_2}$.

\begin{figure}
\includegraphics[width=3in]{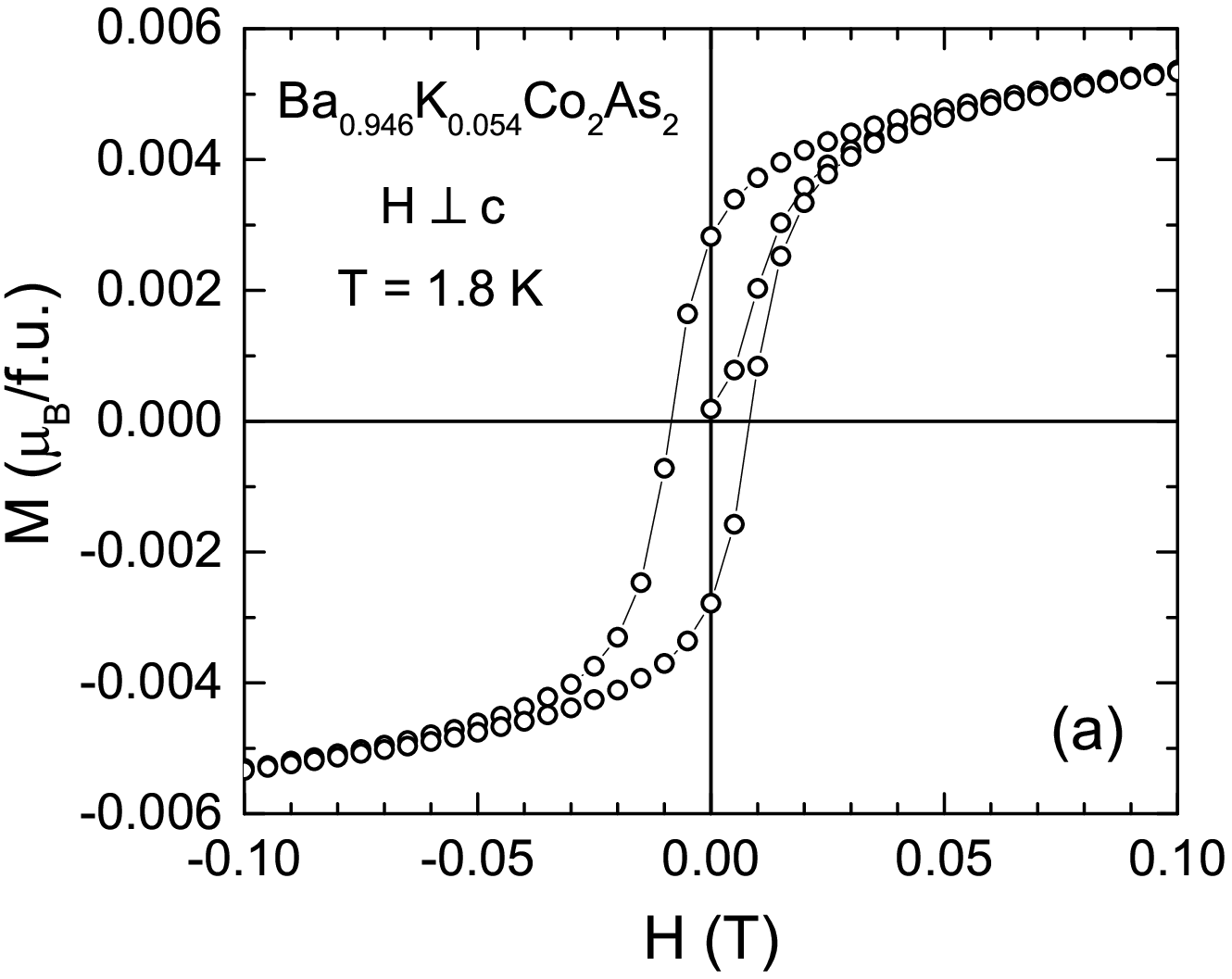}
\includegraphics[width=3in]{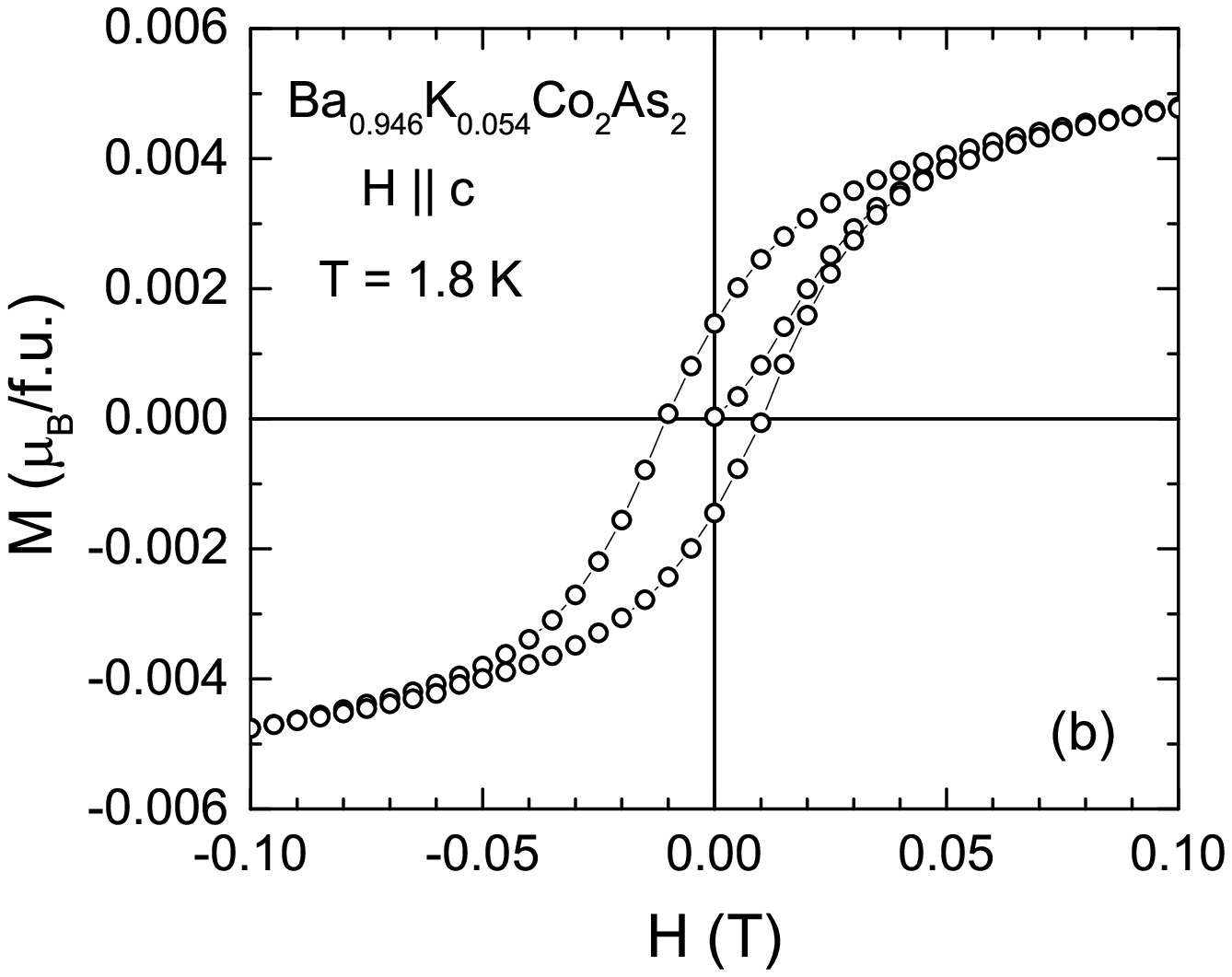}
\caption{Isothermal magnetization $M$ of a ${\rm Ba_{0.946}K_{0.054}Co_2As_2}$ single crystal (self-flux grown) as a function of applied magnetic field $H$ measured at 1.8~K for $H$ applied (a) in the $ab$~plane ($M_{ab}, H \perp  c$) and, (b) along the $c$~axis ($M_c, H \parallel c$).}
\label{fig:MH_BaKCo2As2_lowH}
\end{figure}

Figure~\ref{fig:MH_BaKCo2As2_lowH} shows the hysteresis of $M(H)$ at $T = 1.8$~K on sweeping the field between $\pm 0.10$~T\@.  The hysteresis is typical of a (weak) ferromagnet where here the remanent magnetization is about 0.002--0.003~$\mu_{\rm B}$/f.u.\ for the two field directions and the respective coercive fields are about 0.01~T\@.

\begin{figure}
\includegraphics[width=3in]{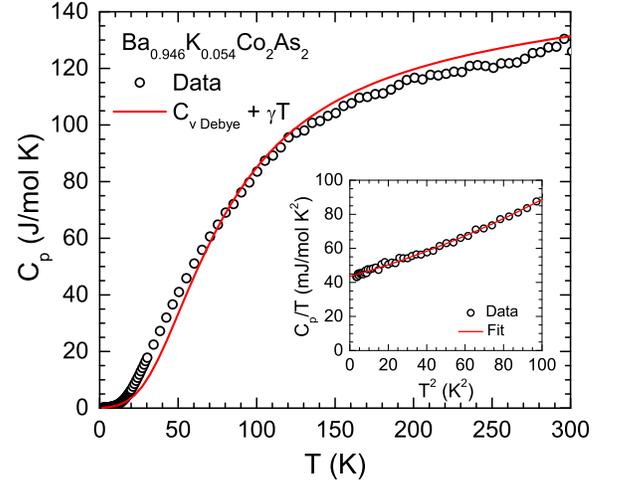}
\caption{(Color online) Heat capacity $C_{\rm p}$ of a ${\rm Ba_{0.946}K_{0.054}Co_2As_2}$ single crystal grown in CoAs~flux as a function of temperature $T$ measured in zero magnetic field. The solid curve is a fit by the sum of the contributions from the Debye lattice heat capacity $C_{\rm V\,Debye}(T)$ and the electronic heat capacity $\gamma T$ according to Eq.~(\ref{eq:Debye_HC-fit}). Inset: $C_{\rm p}/T$ versus $T^2$ below 10~K; the red curve represents the fit of the data by Eq.~(\ref{eq:HC-LowT}) for $1.8~{\rm K} \leq T\leq 10$~K\@.}
\label{fig:HC_BaKCo2As2}
\end{figure}

The $C_{\rm p}(T)$ data of a ${\rm Ba_{0.946}K_{0.054}Co_2As_2}$ crystal are shown in Fig.~\ref{fig:HC_BaKCo2As2}. The low-$T$ $C_{\rm p}(T)$ data do not show any anomaly at or below the onset of the FM-like increase in the low-field $\chi$ at $\sim 18$~K\@. This is likely due to the very small ordered moment.  The low-$T$ $C_{\rm p}(T)$ data for $1.8~{\rm K} \leq T\leq 10$~K shown in the inset of Fig.~\ref{fig:HC_BaKCo2As2} were analyzed by Eq.~(\ref{eq:HC-LowT}) yielding the $\gamma$, $\beta$ and $\delta$ values listed in Table~\ref{tab:tableHC}. The red curve in the inset of Fig.~\ref{fig:HC_BaKCo2As2} shows the fit of $C_{\rm p}(T)/T$. The values of ${\cal D}(E_{\rm F})$ obtained from $\gamma$  and $\Theta_{\rm D}$ obtained from $\beta$ are listed $\Theta_{\rm D}$. Another estimate of $\Theta_{\rm D} = 310(3)$~K is obtained from a fit of the $C_{\rm p}(T)$ data for $1.8~{\rm K} \leq T\leq 300$~K by Eqs.~(\ref{Eqs:AllTCpFit}) which is shown by the red curve in Fig.~\ref{fig:HC_BaKCo2As2}, where we used the analytic Pad\'e approximant fitting function \cite{Ryan2012} for the Debye lattice heat capacity integral.

\begin{figure}
\includegraphics[width=3in]{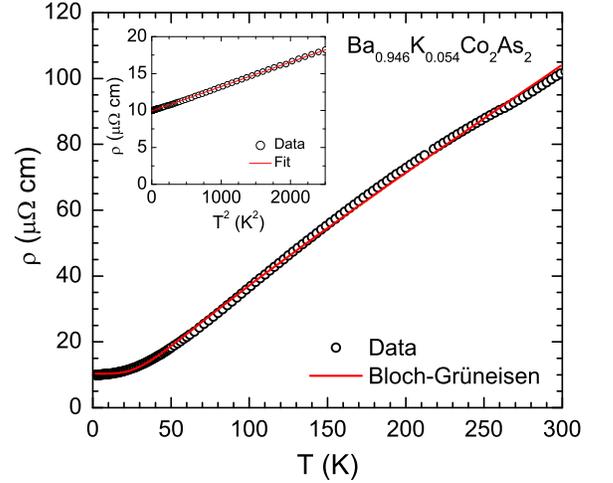}
\caption{(Color online) In-plane electrical resistivity $\rho$ of a ${\rm Ba_{0.946}K_{0.054}Co_2As_2}$ single crystal grown in CoAs~flux as a function of temperature $T$ measured in zero magnetic field. The red solid curve is a fit by the Bloch-Gr\"{u}neisen model. Inset: $\rho$ vs $T^2$ below 50~K\@.   The straight red line is a fit by $\rho = \rho_0+AT^2$ for $1.8~{\rm K} \leq T\leq 50$~K\@.}
\label{fig:rho_BaKCo2As2}
\end{figure}

The in-plane $\rho(T)$ data for a  ${\rm Ba_{0.946}K_{0.054}Co_2As_2}$ crystal are shown in Fig.~\ref{fig:rho_BaKCo2As2}. Like the $C_{\rm p}(T)$ data, the $\rho(T)$ data do not show any anomaly at or below the FM-like onset temperature $\sim 18$~K observed in the $\chi(T)$ data. The $\rho(T)$ is metallic with $\rho(T=1.8\,{\rm K}) = 10.0~\mu \Omega$\,cm and RRR $= 10.2$. The low value of $\rho_0$ and high value of RRR indicate the high quality of the crystal. The $\rho(T)$ data for $1.8~{\rm K} \leq T\leq 50$~K were fitted by Eq.~(\ref{eq:rho_T2}) yielding $\rho_0 = 9.94(1)~\mu \Omega\,{\rm cm}$ and $A = 3.31(1) \times 10^{-3}~\mu \Omega\,{\rm cm/K}^2$. The red curve in the inset of Fig.~\ref{fig:rho_BaKCo2As2} shows the fit of $\rho$ vs $T^2$ data with these parameters. The $T^2$ dependence of $\rho$ at low-$T$ indicates that a Fermi liquid ground state persists in the K-doped crystals. The Kadowaki-Woods ratio $R_{\rm KW} = A/\gamma^2 = 0.17 \times 10^{-5}~\mu \Omega\,{\rm cm/(mJ/mol\,K)^2}$. The $\rho(T)$ data were also fitted by Eqs.~(\ref{Eqs:BGModel}) for 1.8~K~$\leq T \leq$~300~K where again we used the analytic Pad\'e approximant function\cite{Ryan2012} which is shown by the red curve in Fig.~\ref{fig:rho_BaKCo2As2}, yielding $\Theta_{\rm D} = 303(4)$~K\@. The parameters of the various fits to the $C_{\rm p}(T)$ data are summarized in Table~\ref{Tab:RhoFitParams}.

\clearpage

\section{\label{BaKCo2As2_CoAs} Physical Properties of B\lowercase{a}$_{0.935}$K$_{0.065}$C\lowercase{o}$_2$A\lowercase{s}$_2$}

\begin{figure}[b]
\includegraphics[width=3in]{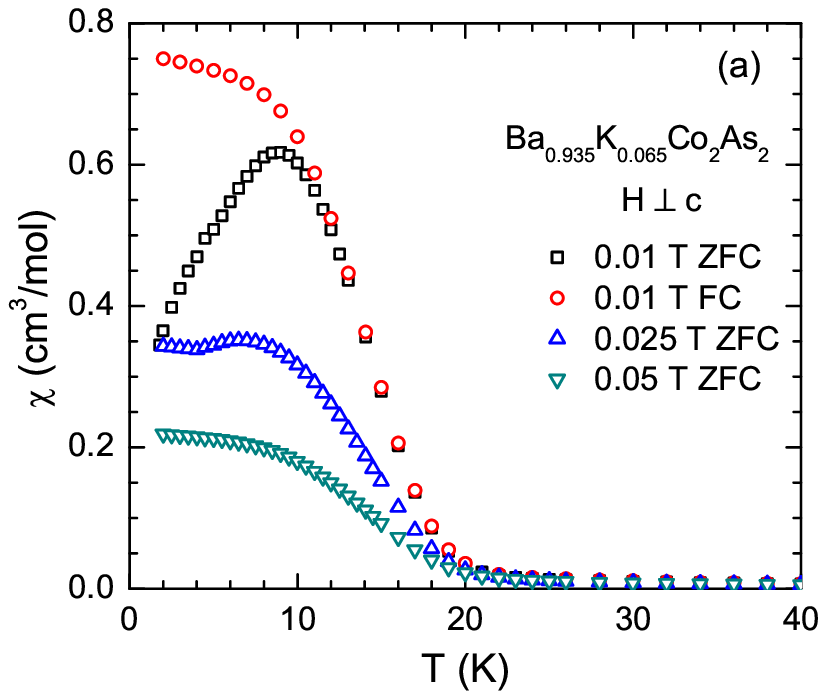}
\includegraphics[width=3in]{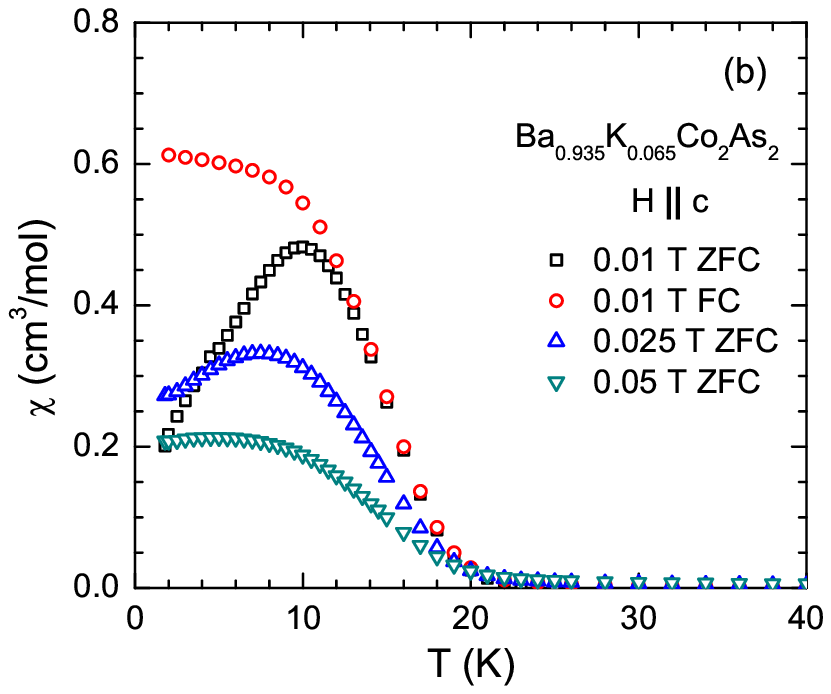}
\caption{(Color online) Zero-field-cooled magnetic susceptibility $\chi$ of a ${\rm Ba_{0.935}K_{0.065}Co_2As_2}$ single crystal grown in CoAs~flux as a function of temperature $T$ in the temperature range 1.8--40~K measured in low magnetic fields $H$ applied (a) in the $ab$~plane ($\chi_{ab}, H \perp  c$) and (b) along the $c$~axis ($\chi_c, H \parallel c$).}
\label{fig:MT_BaKCo2As2_CoAs_Low-H}
\end{figure}

\begin{figure}
\includegraphics[width=3in]{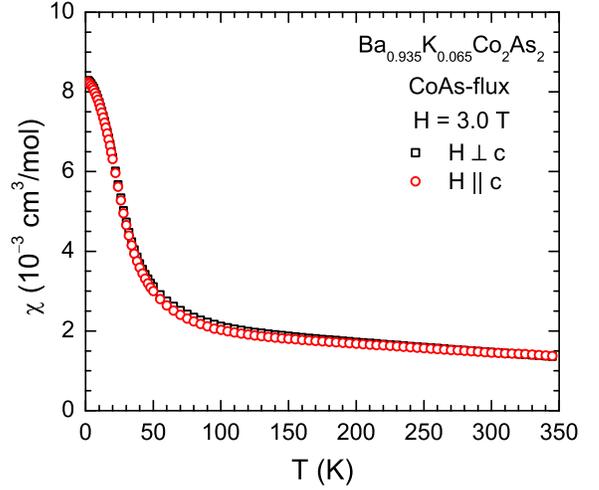}
\caption{(Color online) Zero-field-cooled magnetic susceptibility $\chi$ of a ${\rm Ba_{0.935}K_{0.065}Co_2As_2}$ single crystal grown in CoAs~flux as a function of temperature $T$ in the temperature range 1.8--300~K measured in a magnetic field $H= 3.0$~T applied along the $c$~axis ($\chi_c, H \parallel c$) and in the $ab$~plane ($\chi_{ab}, H \perp  c$).}
\label{fig:MT_BaKCo2As2_CoAs}
\end{figure}

The $\chi(T)$ for a CoAs~flux-grown crystal of ${\rm Ba_{0.935}K_{0.065}Co_2As_2}$ was measured at different values of $H$\@. Here again no superconductivity is observed at $T \geq1.8$~K\@. At $H=0.01$~T the $\chi(T)\equiv M(T)/H$ data shown in Fig.~\ref{fig:MT_BaKCo2As2_CoAs_Low-H} exhibit a rapid FM-like increase below 19~K with an irreversibility between ZFC and FC $\chi$ below 11~K and a peak in the ZFC $\chi$ at 9~K for both $H \parallel c$ and $H \perp  c$. With increasing $H$ the peak becomes broader and $\chi$ becomes almost independent of $T$ below 8~K for $H = 0.05$~T\@. Although the $\chi(T)$ for $H \parallel c$ and $H \perp  c$ exhibit similar behaviors at low~$H$, they have somewhat different magnitudes. On the other hand, from the $\chi(T)$ data measured at $H=3.0$~T that are shown in Fig.~\ref{fig:MT_BaKCo2As2_CoAs} it is evident that there is almost no anisotropy in $\chi$ at this field over the entire $T$-range of measurement.

\begin{figure}
\includegraphics[width=3in]{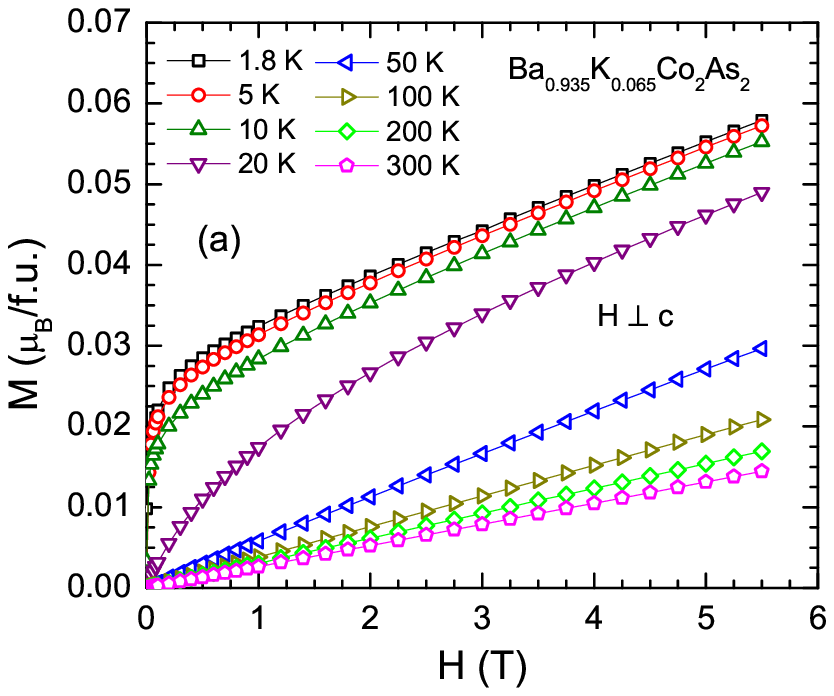}
\includegraphics[width=3in]{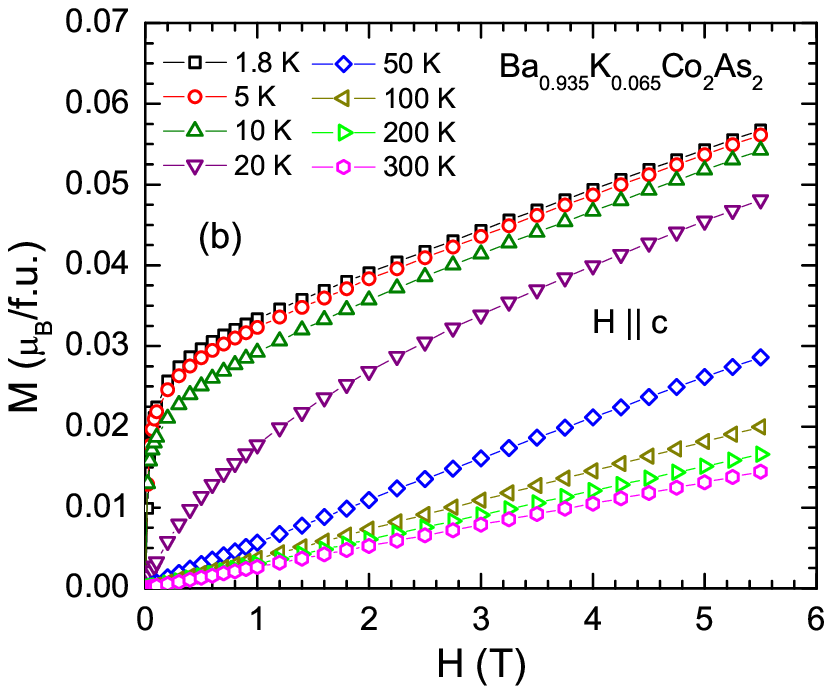}
\caption{(Color online) Magnetization $M$ versus field $H$ of a ${\rm Ba_{0.935}K_{0.065}Co_2As_2}$ crystal grown in CoAs~flux and measured at the indicated temperatures for $H$ applied (a) in the $ab$-plane ($M_{ab}, H \perp  c$) and, (b) along the $c$-axis ($M_c, H \parallel c$).}
\label{fig:MH_BaKCo2As2_CoAs}
\end{figure}

$M(H)$ isotherms for a ${\rm Ba_{0.935}K_{0.065}Co_2As_2}$ single crystal at eight temperatures with $H \perp c$ and $H \parallel c$ are shown in Figs.~\ref{fig:MH_BaKCo2As2_CoAs}(a) and~\ref{fig:MH_BaKCo2As2_CoAs}(b), respectively.  While for $T>50$~K the $M(H)$ isotherms are almost linear in $H$, the low-$T$ $M(H)$ isotherms exhibit a FM-like rapid increase of $M$ with $H$ for $H<0.1$~T\@.  Fitting the high-field data by Eq.~(\ref{eq:MH_linear-fit}), the saturation moments are similar for $H \perp c$ and $H \parallel c$ with the values $M_{\rm s} = 0.028$ and 0.029~$\mu_{\rm B}$/f.u., respectively.  These $M_{\rm s}$ values are about a factor of four larger than obtained from Fig.~\ref{fig:MH_BaKCo2As2} for a self-flux grown crystal of ${\rm Ba_{0.946}K_{0.054}Co_2As_2}$ with a similar composition.

\begin{figure}
\includegraphics[width=3in]{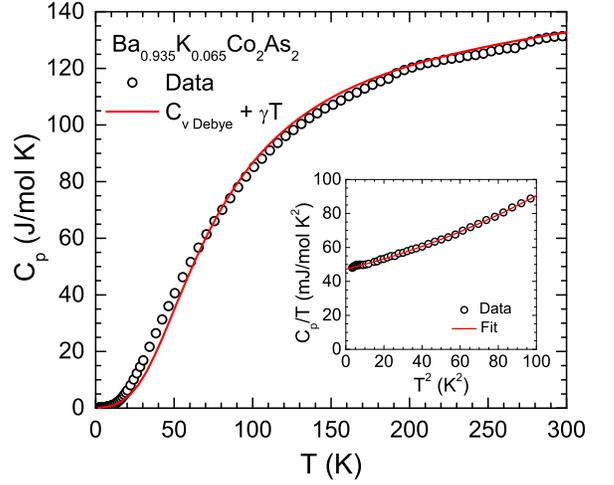}
\caption{(Color online) Heat capacity $C_{\rm p}$ of a ${\rm Ba_{0.935}K_{0.065}Co_2As_2}$ crystal grown in CoAs~flux as a function of temperature $T$ measured in zero magnetic field. The solid curve is the sum of the contributions from the Debye lattice heat capacity $C_{\rm V\,Debye}(T)$ and electronic heat capacity $\gamma T$ according to Eq.~(\ref{eq:Debye_HC-fit}). Inset: $C_{\rm p}/T$ vs $T^2$ plot below 10~K, the red curve represents the fit of $C_{\rm p}/T$ data by Eq.~(\ref{eq:HC-LowT}) for $1.8~{\rm K} \leq T\leq 10$~K\@.}
\label{fig:HC_BaKCo2As2_CoAs}
\end{figure}

The $C_{\rm p}(T)$ data of a ${\rm Ba_{0.935}K_{0.065}Co_2As_2}$ crystal are shown in Fig.~\ref{fig:HC_BaKCo2As2_CoAs}. No anomaly is seen in the low-$T$ $C_{\rm p}(T)$ data would indicate a (weak) ferromagnetic transition, likely due to the very small ordered moment. An analysis of the low-$T$ $C_{\rm p}(T)$ data for $1.8~{\rm K} \leq T\leq 10$~K shown in the inset of Fig.~\ref{fig:HC_BaKCo2As2_CoAs} by Eq.~(\ref{eq:HC-LowT}) gives the fitting parameters $\gamma$, $\beta$ and $\delta$ listed in Table~\ref{tab:tableHC}. The fit is shown by the red curve in the inset of Fig.~\ref{fig:HC_BaKCo2As2_CoAs}. The value of ${\cal D}(E_{\rm F})$ obtained from $\gamma$ and $\Theta_{\rm D}$ obtained from $\beta$ are listed in Table~\ref{tab:tableHC}. A fit of the $C_{\rm p}(T)$ data for $1.8~{\rm K} \leq T\leq 300$~K by Eqs.~(\ref{Eqs:AllTCpFit}) using the analytic Pad\'e approximant fitting function \cite{Ryan2012} for the Debye lattice heat capacity integral gives the value of $\Theta_{\rm D} = 304(3)$ listed in Table~\ref{tab:tableHC}. The fit is shown by the red curve in Fig.~\ref{fig:HC_BaKCo2As2_CoAs}.

\begin{figure}
\includegraphics[width=3in]{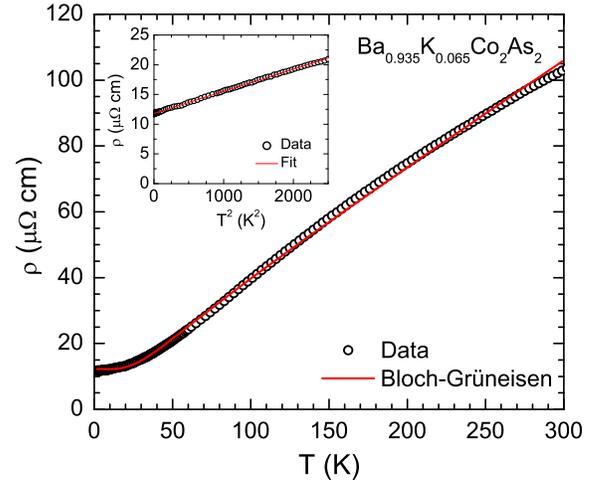}
\caption{(Color online) In-plane electrical resistivity $\rho$ of a ${\rm Ba_{0.935}K_{0.065}Co_2As_2}$ crystal grown in CoAs~flux as a function of temperature $T$ measured in zero magnetic field. The red solid curves are the fit by the Bloch-Gr\"{u}neisen model. Inset: $\rho$ vs $T^2$ plots below 50~K, the solid line is the fit by $\rho = \rho_0+AT^2$ for $1.8~{\rm K} \leq T\leq 50$~K.}
\label{fig:rho_BaKCo2As2_CoAs}
\end{figure}

The in-plane $\rho(T)$ data of a ${\rm Ba_{0.935}K_{0.065}Co_2As_2}$ crystal are shown in Fig.~\ref{fig:rho_BaKCo2As2_CoAs}. A metallic behavior is inferred from the $T$ dependence of $\rho$ without any anomaly in low-$T$ data. The residual resistivity is $\rho(T=1.8~{\rm K}) = 11.8~\mu \Omega$\,cm and RRR $\approx 9$, demonstrating the high quality of the crystal. An analysis of the $\rho(T)$ data for $1.8~{\rm K} \leq T\leq 50$~K by Eq.~(\ref{eq:rho_T2}) gives $\rho_0 = 11.81(2)~\mu \Omega\,{\rm cm}$ and $A = 3.72 \times 10^{-3}~\mu \Omega\,{\rm cm/K}^2$. The fit is shown by the red curve in the inset of Fig.~\ref{fig:rho_BaKCo2As2_CoAs}. The Kadowaki-Woods ratio $R_{\rm KW} = A/\gamma^2 = 0.17 \times 10^{-5}~\mu \Omega\,{\rm cm/(mJ/mol\,K^2)}$. The $\rho(T)$ data were fitted by Eqs.~(\ref{Eqs:BGModel}) for 1.8~K~$\leq T \leq$~300~K using the analytic Pad\'e approximant function\cite{Ryan2012} as shown by red curve in Fig.~\ref{fig:rho_BaKCo2As2_CoAs}. The fit parameters are summarized in Table~\ref{Tab:RhoFitParams}.

\clearpage

\section{\label{BaKCo2As2_Sn} Physical Properties of B\lowercase{a}$_{0.78}$K$_{0.22}$C\lowercase{o}$_2$A\lowercase{s}$_2$}

\begin{figure}[b]
\includegraphics[width=3in]{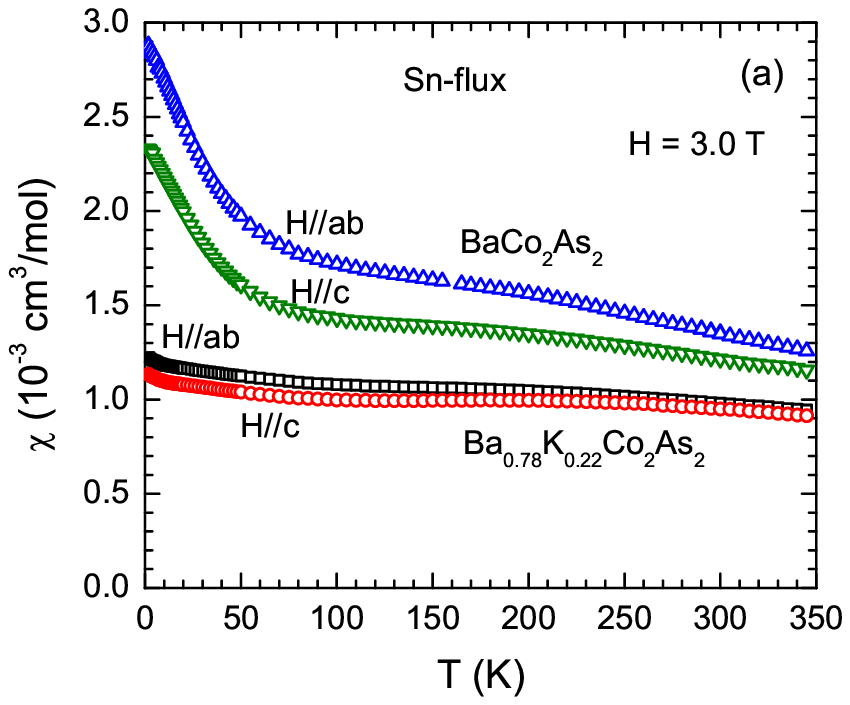}
\includegraphics[width=3in]{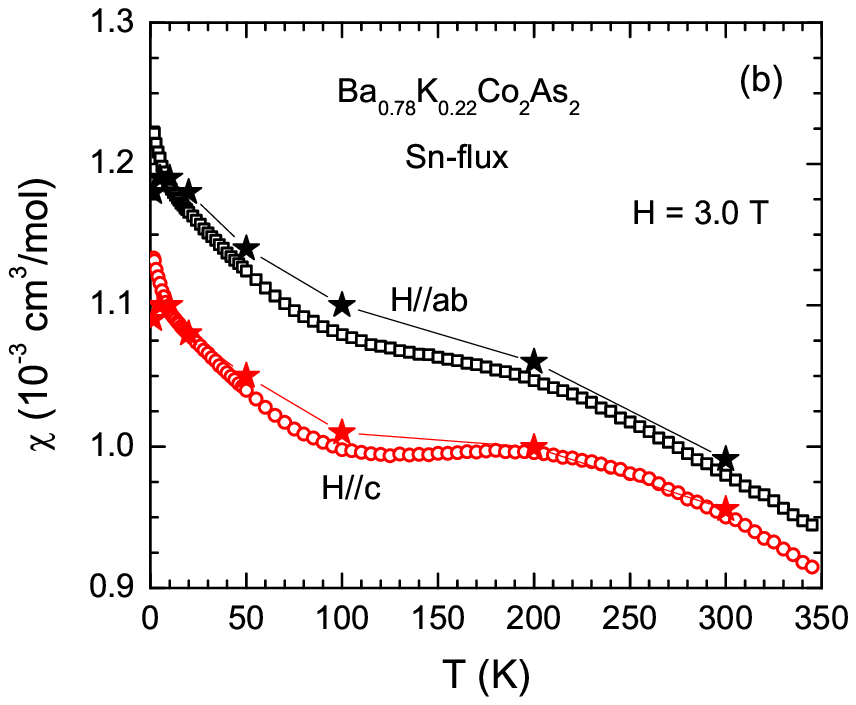}
\caption{(Color online) (a) Zero-field-cooled magnetic susceptibility $\chi$ of a ${\rm Ba_{0.78}K_{0.22}Co_2As_2}$ single crystal grown in Sn~flux versus temperature $T$ in the $T$ range 1.8--300~K measured in a magnetic field $H= 3.0$~T applied along the $c$~axis ($\chi_c, H \parallel c$) and in the $ab$~plane ($\chi_{ab}, H \perp  c$). Also shown are the $\chi(T)$ data of a Sn-flux grown ${\rm BaCo_2As_2}$ crystal from Fig.~\ref{fig:MT_BaCo2As2}. (b) The $\chi(T)$ of ${\rm Ba_{0.78}K_{0.22}Co_2As_2}$ on an expanded scale. The stars of respective colors represent the $\chi$ obtained from fitting $M(H)$ isotherm data by Eq.~(\ref{eq:MH_linear-fit}). The solid lines connecting the stars are guides to eye.}
\label{fig:MT_BaKCo2As2_Sn}
\end{figure}

The ${\rm Ba_{0.78}K_{0.22}Co_2As_2}$ crystals discussed in this section were grown in Sn flux.

The $\chi(T) \equiv M(T)/H$ data of a ${\rm Ba_{0.78}K_{0.22}Co_2As_2}$ crystal measured in $H = 3.0$~T are shown in Fig.~\ref{fig:MT_BaKCo2As2_Sn}. Here again no superconductivity is observed at $T \geq1.8$~K\@. A comparison of the $\chi$ of ${\rm Ba_{0.78}K_{0.22}Co_2As_2}$ with that of Sn-flux grown ${\rm BaCo_2As_2}$ is shown in Fig.~\ref{fig:MT_BaKCo2As2_Sn}(a). It is seen that the magnitude of $\chi$ decreases with K-doping. An expanded view of $\chi(T)$ of ${\rm Ba_{0.78}K_{0.22}Co_2As_2}$ is shown in Fig.~\ref{fig:MT_BaKCo2As2_Sn}(b) which shows a broad peak in $\chi$ near 100~K\@. A similar feature has been observed in $T$ dependence of $\chi$ of ${\rm SrCo_2As_2}$. \cite{Pandey2013b}

\begin{figure}
\includegraphics[width=3in]{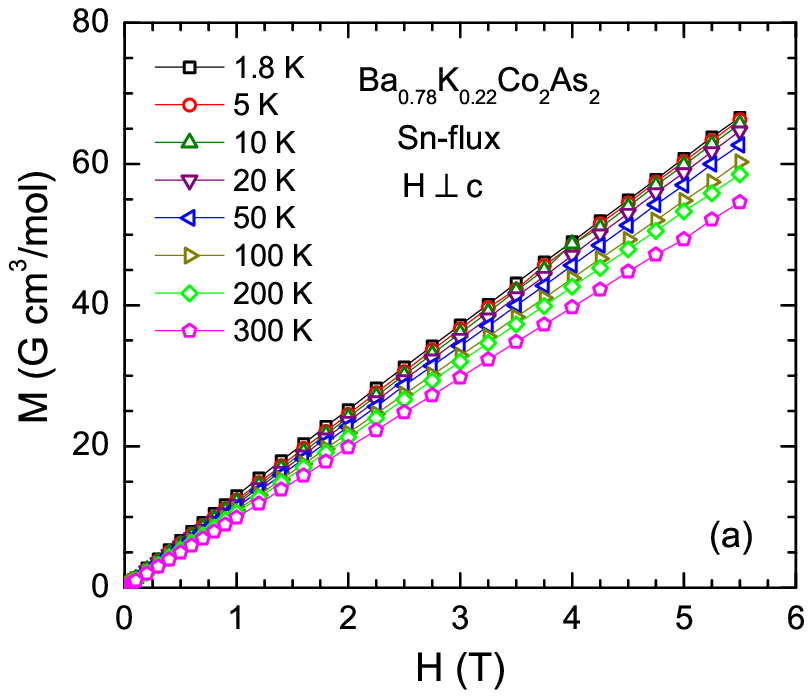}
\includegraphics[width=3in]{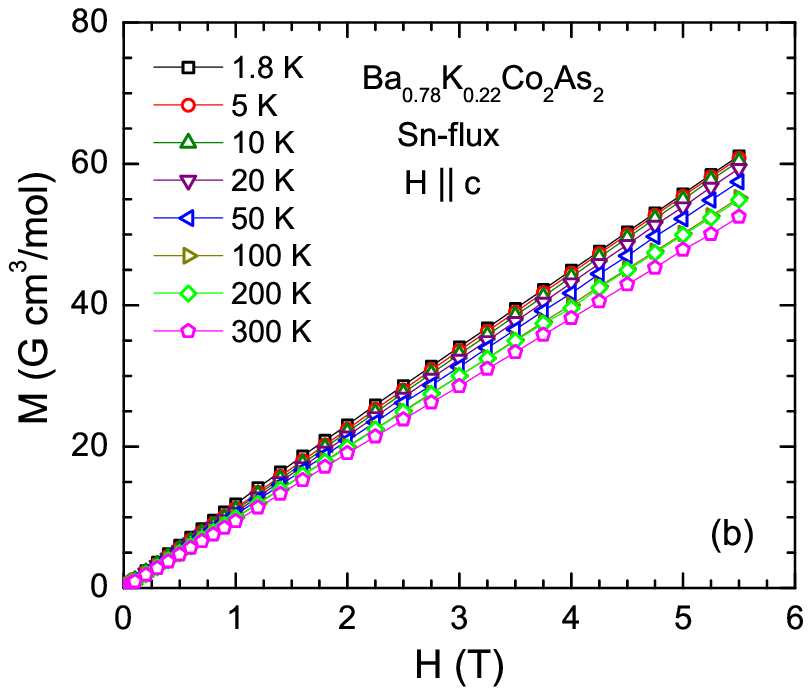}
\caption{(Color online) Isothermal magnetization $M$ of a ${\rm Ba_{0.78}K_{0.22}Co_2As_2}$ crystal grown in Sn~flux as a function of applied magnetic field $H$ measured at the indicated temperatures for $H$ applied (a) in the $ab$-plane ($M_{ab}, H \perp  c$) and, (b) along the $c$-axis ($M_c, H \parallel c$).}
\label{fig:MH_BaKCo2As2_Sn}
\end{figure}

The $M(H)$ isotherms of a ${\rm Ba_{0.78}K_{0.22}Co_2As_2}$ crystal at eight temperatures are shown in Fig.~\ref{fig:MH_BaKCo2As2_Sn}. These are similar to those of a Sn~flux-grown ${\rm BaCo_2As_2}$ crystal shown in Fig.~\ref{fig:MH_BaCo2As2_Sn}.  In particular, the $M(H)$ isotherms are almost linear in $H$ at all temperatures. The magnetic behavior of this ${\rm Ba_{0.78}K_{0.22}Co_2As_2}$ crystal is therefore much different from those of ${\rm Ba_{0.946}K_{0.054}Co_2As_2}$ and ${\rm Ba_{0.935}K_{0.065}Co_2As_2}$ grown out of CoAs~flux discussed above. In particular, no FM-like behaviors are observed in the $M(H)$ isotherms in Fig.~\ref{fig:MH_BaKCo2As2_Sn}.

\begin{figure}
\includegraphics[width=3in]{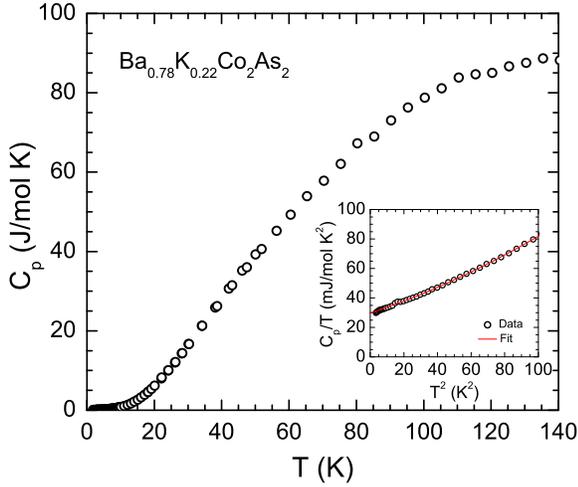}
\caption{(Color online) Heat capacity $C_{\rm p}$ of a ${\rm Ba_{0.78}K_{0.22}Co_2As_2}$ crystal grown in Sn~flux as a function of temperature $T$ measured in zero magnetic field. Inset: $C_{\rm p}/T$ versus $T^2$ below 10~K;  the red curve represents the fit of $C_{\rm p}/T$ data by Eq.~(\ref{eq:HC-LowT}) for $4~{\rm K} \leq T\leq 10$~K\@.}
\label{fig:HC_BaKCo2As2_Sn}
\end{figure}

The $C_{\rm p}(T)$ data of a ${\rm Ba_{0.78}K_{0.22}Co_2As_2}$ crystal are shown in Fig.~\ref{fig:HC_BaKCo2As2_Sn}. No anomaly is observed in $C_{\rm p}(T)$ except for an extrinsic weak kink near 3.8~K due to the superconductivity of a very small amount of Sn flux attached to the crystal. A fit of the low-$T$ $C_{\rm p}(T)/T$ versus $T^2$ data in the inset of Fig.~\ref{fig:HC_BaKCo2As2_CoAs} for $4.0~{\rm K} \leq T\leq 10$~K by Eq.~(\ref{eq:HC-LowT}) gives the $\gamma$, $\beta$ and $\delta$ values listed in Table~\ref{tab:tableHC}. The fit is shown as the red curve in the inset of Fig.~\ref{fig:HC_BaKCo2As2_CoAs}. The values of ${\cal D}(E_{\rm F})$ obtained from $\gamma$ and $\Theta_{\rm D}$ obtained from $\beta$ are also listed in Table~\ref{tab:tableHC}.

\begin{figure}
\includegraphics[width=3in]{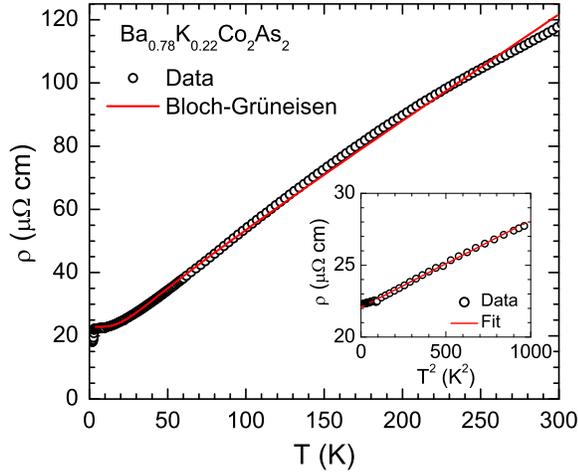}
\caption{(Color online) In-plane electrical resistivity $\rho$ of a ${\rm Ba_{0.78}K_{0.22}Co_2As_2}$ single crystal grown in Sn~flux as a function of temperature $T$ measured in zero magnetic field. The red solid curve is  the fit by the Bloch-Gr\"{u}neisen model. Inset: $\rho$ vs $T^2$ for $T<32$~K; the straight red line is a fit of the data by $\rho = \rho_0+AT^2$ for $10.0~{\rm K} \leq T\leq 30$~K.}
\label{fig:rho_BaKCo2As2_Sn}
\end{figure}

The in-plane $\rho(T)$ data of a ${\rm Ba_{0.78}K_{0.22}Co_2As_2}$ crystal in $H=0$ are shown in Fig.~\ref{fig:rho_BaKCo2As2_Sn}. A metallic behavior is demonstrated by the magnitude and $T$ dependence of $\rho$. A sharp drop in $\rho$ below $\sim 3.8$~K is due to the superconductivity of a small amount of Sn flux attached to the crystal. No other anomaly is seen in the $\rho(T)$ data. The residual resistivity  $\rho(T=3.8~{\rm K}) = 22.3~\mu \Omega$\,cm and RRR $= 5.4$. A fit of the $\rho$ versus $T^2$ data in the inset of Fig.~\ref{fig:rho_BaKCo2As2_Sn} for $10~{\rm K} \leq T\leq 30$~K by Eq.~(\ref{eq:rho_T2}) gives $\rho_0 = 22.07(3)~\mu \Omega\,{\rm cm}$ and $A = 6.06(5) \times 10^{-3}~\mu \Omega\,{\rm cm/K}^2$. The fit is shown as the red straight line in the inset.  The Kadowaki-Woods ratio $R_{\rm KW} = A/\gamma^2 = 0.69 \times 10^{-5}~\mu \Omega\,{\rm cm/(mJ/mol\,K^2)}$. The $\rho(T)$ data were further fitted by the Bloch-Gr\"uneisen model in Eqs.~(\ref{Eqs:BGModel}) for 4.0~K~$\leq T \leq$~300~K using the analytic Pad\'e approximant function\cite{Ryan2012} as shown by the red curve in Fig.~\ref{fig:rho_BaKCo2As2_CoAs}.  The fit parameters are summarized in Table~\ref{Tab:RhoFitParams}.

\section{\label{Conclusion} Summary}

The powder x-ray and neutron diffraction studies of ${\rm BaCo_2As_2}$, ${\rm Ba_{0.94}K_{0.06}Co_2As_2}$ and ${\rm Ba_{0.78}K_{0.22}Co_2As_2}$ at room temperature showed that the three compositions crystallize in the same body-centered tetragonal ${\rm ThCr_2Si_2}$-type structure.  Additional neutron powder diffraction data for ${\rm BaCo_2As_2}$ were obtained down to 10~K, which revealed no evidence for temperature-induced lattice distortions.  On the other hand, these data revealed that the tetragonal $c$~axis lattice parameter shows negative thermal expansion over this $T$ range, whereas the in-plane $a$~axis parameter and the unit cell volume show normal expansion on increasing the temperature.  The $T$-dependent behaviors of $a$ and~$c$ are similar to those found previously for ${\rm SrCo_2As_2}$.\cite{Pandey2013b}

The electrical resistivity data for all three compounds exhibit metallic character and a $T^2$ temperature dependence at low temperatures indicative of a Fermi-liquid ground state.  The heat capacity and electrical resistivity data show no evidence for any phase transitions between 2 and~300~K for ${\rm BaCo_2As_2}$.  Our magnetization and magnetic susceptibility data for crystals from two batches of  ${\rm Ba_{0.94}K_{0.06}Co_2As_2}$ crystals grown in CoAs self-flux showed weak ferromagnetism below $\sim 10$~K with small saturation moments of $\approx 0.007$ and 0.03~${\rm \mu_{\rm B}}$/f.u., respectively, whereas a crystal of ${\rm Ba_{0.78}K_{0.22}Co_2As_2}$ grown in Sn flux showed no such behavior.  This nonmonotonic behavior of the weak ferromagnetism  versus K content may be associated with the different fluxes used to grow the crystals with the 6\% and 22\% potassium concentrations.  No evidence for bulk superconductivity was found for any of the crystals above 1.8~K.

The low-temperature heat capacity data for the three compounds show similar and rather large Sommerfeld electronic heat capacity coefficients, indicating large densities of states at the Fermi energy ${\cal D}(E_{\rm F})$.  We infer that the factor of two enhancement of this ${\cal D}(E_{\rm F})$ above the band structure value arises from a combination of many-body electron-electron effects that renormalize the band structure (decreasing the band widths) and the effect of the electron-phonon interaction.  For ${\rm BaCo_2As_2}$, the former enhancement was previously estimated to be $m^\ast/m_{\rm band} \approx 1.4$,\cite{Xu2013} which we use to compute the latter contribution to be $\approx 1.6$ that in turn yields the electron-phonon coupling constant $\lambda_{\rm el-ph} \approx 0.6$.  This value of $\lambda_{\rm el-ph}$ is similar to the value of 0.76 obtained from {\it ab initio} calculations for ${\rm BaNi_2As_2}$.\cite{Subedi2008}

Our band structure value for the density of states at the Fermi energy for ${\rm BaCo_2As_2}$, ${\cal D}_{\rm band}(E_{\rm F}) = 8.23$~states/(eV\,f.u.) for both spin directions, is similar to the value of 8.5~states/(eV\,f.u.) for both spin directions found in Ref.~\onlinecite{Sefat2009}.  These values are sufficiently large that on the basis of an itinerant magnetism picture one expects ${\rm BaCo_2As_2}$ to be an itinerant ferromagnet,\cite{Sefat2009} which is not observed.  To explain the lack of ferromagnetism, Sefat et al.\ suggested that ${\rm BaCo_2As_2}$ is close to a quantum critical point where quantum critical fluctuations destabilize long-range ferromagnetic order.\cite{Sefat2009}

The $T$ dependences of our intrinsic $\chi$ and $^{75}$As NMR shift data for CoAs~flux-grown single-crystals of ${\rm BaCo_2As_2}$ are the same, indicating that the $T$ dependences of $\chi_{ab}$ and $\chi_c$ are intrinsic to ${\rm BaCo_2As_2}$.  These $T$~dependences are similar to those reported previously for $\chi(T)$ of a ${\rm BaCo_2As_2}$ crystal by Sefat et al.,\cite{Sefat2009} but their magnitude of $\chi$ is much larger than ours and they reported an anisotropy with $\chi_{ab} < \chi_{c}$ that is opposite to ours.  As discussed in Ref.~63 of Ref.~\onlinecite{Pandey2013b}, these differences between our $\chi(T)$ results and those of Ref.~\onlinecite{Sefat2009} likely result from the presence of ferromagnetic impurities in their crystal combined with  their low measurement field of 0.1~T\@.

We find no clear evidence from our $M(H)$ isotherms or our $\chi(T)$ data that ${\rm BaCo_2As_2}$ is near a ferromagnetic quantum critical point as suggested in Ref.~\onlinecite{Sefat2009}.  The $^{75}$As NMR dynamics we observe can be explained by stripe-type AFM fluctuations, FM fluctuations or a combination of these.  If a material is near a quantum critical point, one might expect that the properties are sensitive to doping.  However, our crystals of K-doped ${\rm Ba_{0.78}K_{0.22}Co_2As_2}$ grown in Sn~flux exhibit magnetic properties similar to those of undoped ${\rm BaCo_2As_2}$.  On the other hand, two batches of ${\rm Ba_{0.94}K_{0.0.06}Co_2As_2}$ crystals grown in CoAs self-flux show weak ferromagnetism with small ordered moments of $\approx 0.007$ and $0.03~\mu_{\rm B}$/f.u., respectively.

Interestingly, this conundrum between ferromagnetic and antiferromagnetic interactions and ordering was present in the early days of the iron-arsenide high-$T_{\rm c}$ field in 2008, where the band structure density of states ${\cal D}_{\rm band}(E_{\rm F})$ calculated for the parent LaFeAsO compound was approximately large enough to lead to itinerant ferromagnetism,\cite{Singh2008} whereas itinerant antiferromagnetism (a spin-density wave) was observed instead,\cite{Johnston2010} indicating a competition between these magnetic structures.  A related observation is that ${\rm SrCo_2As_2}$ shows no long-range magnetic ordering but exhibits a broad maximum in $\chi(T)$ at about 115~K typically associated with low-dimensional dynamic antiferromagnetic correlations.\cite{Pandey2013b}  Furthermore, from inelastic neutron scattering measurements ${\rm SrCo_2As_2}$ shows rather strong antiferromagnetic fluctuations at the same stripe wave vector at which the FeAs-based materials exhibit AFM fluctuations and long-range antiferromagnetic ordering,\cite{Jayasekara2013} whereas the calculated ${\cal D}_{\rm band}(E_{\rm F})$ suggests incipient ferromagnetism.\cite{Pandey2013b}  On the other hand, the isostructural compound ${\rm CaCo_{1.86}As_2}$ exhibits dominant ferromagnetic  correlations,\cite{Cheng2012, Ying2012, Anand2014} and the magnetically ordered state consists of ferromagnetically aligned layers of Co spins in the $ab$~plane that are antiferromagnetically aligned from layer to layer along the $c$~axis to form an A-type antiferromagnetic structure.\cite{Quirinale2013}  It will be interesting to resolve the puzzle of how these conflicting interactions and magnetic structures compete and compromise in the $A{\rm Co_2As_2}$ compounds.

\acknowledgments

We thank Abhishek Pandey for experimental assistance and for many helpful discussions. The research at Ames Laboratory was supported by the U.S. Department of Energy, Office of Basic Energy Sciences, Division of Materials Sciences and Engineering. Ames Laboratory is operated for the U.S. Department of Energy by Iowa State University under Contract No.~DE-AC02-07CH11358.  V.V.O.\ thanks the Ames Laboratory-USDOE for providing the opportunity to be a visiting scientist at the Laboratory and also thanks the Russian Foundation for Basic Research (No.~12-02-31814) for support.  Use of the National Synchrotron Light Source, Brookhaven National Laboratory, was supported by the U.S. Department of Energy, Office of Basic Energy Sciences, under Contract No.~DE-AC02-98CH10886. Research at the Spallation Neutron Source at Oak Ridge National Laboratory was sponsored by the Scientific User Facilities Division, Office of Basic Energy Sciences, U.S. Department of Energy.

\end{document}